\documentclass[preprint]{iucr}
\journalcode{JSR}

\usepackage{amsmath}
\usepackage{bm}
\usepackage{amssymb}
\usepackage{dcolumn}
\usepackage[acronym]{glossaries}
\usepackage{graphicx}
\usepackage{lineno}
\usepackage{siunitx}
\usepackage{miller}
\usepackage[version=4]{mhchem}
\usepackage{tabularx}
\usepackage{multirow}
\usepackage[dvipsnames]{xcolor}
\usepackage[colorlinks]{hyperref}
\hypersetup{
    linkcolor=Black,
    citecolor=Black,
    filecolor=Black,
    urlcolor=Black,}

\setacronymstyle{long-short}
\newacronym{mc}{MC}{Monte Carlo}
\newacronym{sh}{SH}{Spherical Harmonic}
\newacronym{saxs}{SAXS}{Small-angle X-ray Scattering}
\newacronym{saxstt}{SAXSTT}{Small-angle X-ray Scattering Tensor Tomography}
\newacronym{sigtt}{SIGTT}{Spherical Integral Geometric Tensor Tomography}
\newacronym{rsm}{RSM}{reciprocal space map}
\newacronym{pmma}{PMMA}{polymethyl methylacrylate}

\newcommand{\chalmersphys}{
    Department of Physics,
    Chalmers University of Technology,
    Gothenburg, Sweden}
\newcommand{\affilpsi}{
    Photon Science Division,
    Paul Scherrer Institute (PSI),
    Villigen,
    Switzerland}
\newcommand{\affilepfl}{
    Institute of Materials,
    École Polytechnique Fédérale de Lausanne (EPFL),
    Lausanne,
    Switzerland}  
\makeatletter

\providecommand\HyperFirstAtBeginDocument{\AtBeginDocument}
\HyperFirstAtBeginDocument{\ifx\hyper@anchor\@undefined
\global\let\oldnewlabel\newlabel
\gdef\newlabel#1#2{\newlabelxx{#1}#2}
\gdef\newlabelxx#1#2#3#4#5#6{\oldnewlabel{#1}{{#2}{#3}}}
\AtEndDocument{\ifx\hyper@anchor\@undefined
\let\newlabel\oldnewlabel
\fi}
\fi}
\begin{document}

\title{Investigating the missing wedge problem in small-angle x-ray scattering tensor tomography across real and reciprocal space}

\aff[a]{\chalmersphys}
\aff[b]{\affilpsi}
\aff[c]{\affilepfl}
\author[a]{Leonard C.}{Nielsen}
\author[b,c]{Torne}{Tänzer}
\author[b]{Irene}{Rodriguez-Fernandez}
\author[a]{Paul}{Erhart}
\cauthor[a,b,c]{Marianne}{Liebi}{marianne.liebi@psi.ch}

\begin{abstract}
Small-angle scattering tensor tomography is a technique for studying anisotropic nanostructures of millimeter-sized samples in a volume-resolved manner.
It requires the acquisition of data through repeated tomographic rotations about an axis which is subjected to a series of tilts.
The tilt that can be achieved with a typical setup is geometrically constrained, which leads to limits in the set of directions from which the different parts of the \gls{rsm} can be probed.
Here, we characterize the impact of this limitation on reconstructions in terms of the missing wedge problem of tomography, by treating the problem of tensor tomography as the reconstruction of a three-dimensional field of functions on the unit sphere, represented by a grid of Gaussian radial basis functions.
We then devise an acquisition scheme to obtain complete data by remounting the sample, which we apply to a sample of human trabecular bone.
Performing tensor tomographic reconstructions of limited data sets as well as the complete data set, we further investigate and validate the missing wedge understanding of data incompleteness by investigating reconstruction errors due to data incompleteness across both real and reciprocal space.
Finally, we carry out an analysis of orientations and derived scalar quantities, to quantify the impact of this missing wedge problem on a typical tensor tomographic analysis.
We conclude that the effects of data incompleteness are consistent with the predicted impact of the missing wedge problem, and that the impact on tensor tomographic analysis is appreciable but limited, especially if precautions are taken.
In particular, there is only limited impact on the means and relative anisotropies of the reconstructed reciprocal space maps.
\end{abstract}

\maketitle

\section{Introduction}



\Gls{saxstt} is a promising method for probing anisotropic nanostructures of macroscopic samples in a volume-resolved manner \cite{liebi_nat_2015, schaff_nature_2015, liebi_aca_2018, manuel_jsr_2020}.
It has been applied to the study of a variety of biological materials, including bone, tendon, and myelin \cite{marios_2021, casanova_2023_biomaterials, grunewald_2023_iucrj, silvabaretto_acta_2024}.
In the absence of very strong real-space uniformity and reciprocal-space symmetry constraints \cite{stribeck_mcp_2006, skjonsfjell_jac_2016}, \gls{saxstt} requires a more general acquisition scheme than traditional scalar tomography, such as rotating the sample while subjecting the axis of rotation to a series of tilts \cite{schaff_nature_2015, liebi_nat_2015, liebi_aca_2018}, carrying out measurements over part of a sphere of rotation.
Such acquisition schemes are generally geometrically constrained to a tilt of up to \qty{45}{\degree}, since the rotation stage will obstruct the beam at greater tilt angles.
In \citeasnoun{nielsen_tt_2023} using this measurement scheme with simulated data, it was observed that the degree of correlation with the original \glspl{rsm} approached lower values than what should be theoretically attainable in terms of the \gls{rsm} representations used in the simulations and reconstructions, even at very low noise levels.
This can likely be attributed in part to the so-called missing wedge problem, a common data incompleteness problem in tomography \citeaffixed{liu_missingwedge_2018}{\textit{e.g.}, }.
While a limited investigation into the effect of reduced data was carried out in \citeasnoun{liebi_aca_2018}, a thorough examination of the effects of data incompleteness is still outstanding.
A deeper understanding of data incompleteness in \gls{saxstt} is a crucial component of the development of approaches to counteract this incompleteness, similar to those used in other tomography methods, as in, e.g., \citeasnoun{trampert_missing_wedge}, \citeasnoun{ding_missing_wedge}, and \citeasnoun{moebel_missing_wedge}.

Here, to investigate these effects under real experimental conditions we present a scheme utilizing sample remounting to yield two incomplete data sets, each measured using the \qtyrange[range-units=repeat]{0}{45}{\degree} tilt scheme, which when combined form a complete data set.
The scheme was applied in measurements on a sample of human trabecular bone.
For the reconstruction, we employed a \gls{rsm} representation which uses local Gaussian radial basis functions on a spherical grid to interpolate measured data.
This reconstruction is closely related to the \gls{sigtt} approach \cite{nielsen_tt_2023} but replaces the model for the \gls{rsm} with local functions, which avoids artefacts due to the spherical harmonic Gibbs phenomenon \cite{gelb_1997_gibbs}.
Both models have in common that the only symmetry enforced is Friedel symmetry, and thus allow for reconstruction of complex textures.
In addition, the use of local radial basis functions permits the problem of \gls{saxstt} to be analyzed as a set of scalar tomography problems, which allows the application of the framework of standard tomographic analysis.
By carrying out reconstructions from the two separate data sets, as well as the combined data set, and comparing the reconstructions, this work seeks to investigate whether imperfect reconstructions in limited-angle \gls{saxstt} can indeed be attributed to missing wedges.
In addition, we aim to provide insight into the impact of this effect on \gls{saxstt} analysis.
We conclude that the differences between complete and partial data sets are consistent with the predicted effects of the missing wedge problem.
These effects impact typical \gls{saxstt} analysis in a non-trivial but manageable way, and suggest strategies for mitigation.
In particular, two important scalar quantities, the means and relative anisotropies of the reconstructed reciprocal space maps, show relatively little impact from the missing wedges.

\section{Theory}
\label{sect:Theory}
\begin{figure}
    \includegraphics[width=\linewidth]{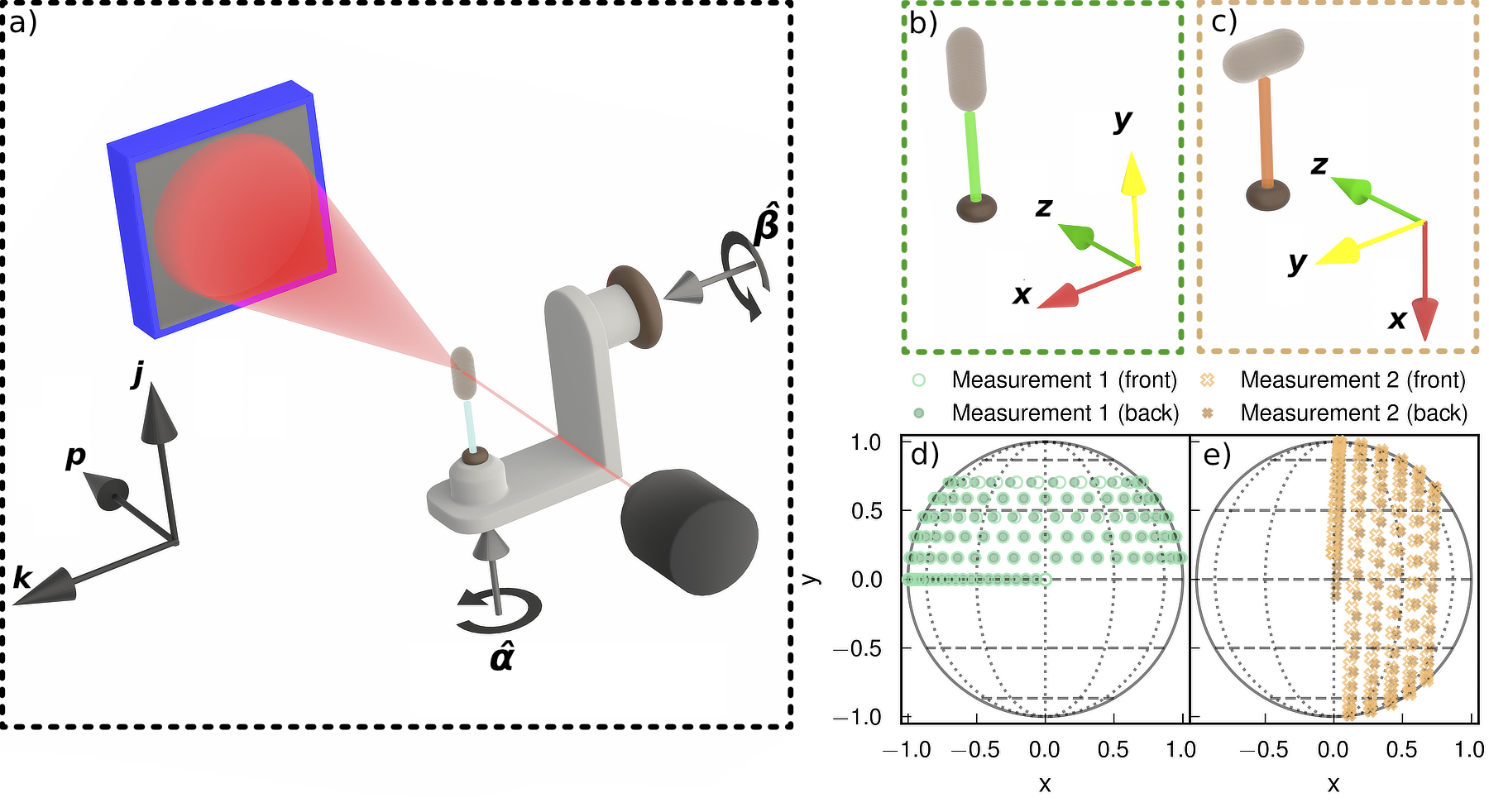}
  \caption{
    \textbf{\Gls{saxstt} measurement setup and probed directions.}
    \textbf{a)} Full setup with detector, goniometer with sample mounted on a pin, x-ray source, and measurement coordinate system.
    \textbf{b)} Initial sample mounting during first measurement and initial orientation of sample coordinate system.
    \textbf{c)} Initial sample mounting during second measurement.
    \textbf{d)} Points on the sphere of projection probed during first measurement.
    \textbf{e)} Points probed during second measurement.
    The points on the sphere of projection give the coordinates of the projection direction $\bm{p}$ in the sample coordinate system spanned by $\bm{x}$, $\bm{y}$ and $\bm{z}$.
    }
  \label{fig:sphere_of_proj}
\end{figure}

The \gls{rsm} measured by \gls{saxs} in a small volume may be written as
\begin{linenomath*}\begin{align}
    \textsc{rsm}(\bm{q}, \bm{r})
    &=
    \iiint dV \big[\Tilde{\rho}(\bm{r}' - \bm{r}) \exp(-i \bm{q} \cdot (\bm{r}' - \bm{r}))],
    \label{eq:reciprocal_space_map}
\end{align}\end{linenomath*}
where $\tilde{\rho}(\bm{r}' - \bm{r})$ is the auto-correlation function of the electron density over the small volume, $\bm{r}$ is the position of the center of the volume, $\bm{r'}$ is the point of integration within the volume, and $\bm{q}$ is the reciprocal space vector.
For a small scattering angle, such that the scattered intensity travels approximately the same path as the transmitted intensity, and assuming that the total amount of scattering is small enough to not significantly influence the transmission, we can probe the \gls{rsm} by measuring the small-angle scattering intensity with a small beam, and correcting by the transmitted intensity as
\begin{linenomath*}\begin{align}
    \int_{p_0}^{p_\text{end}}\mathop{dp}\textsc{rsm}(\bm{q}, j, k, p)
    &\propto
    \frac{I_S(\bm{q}, j, k)}{I_T(j, k)},
    \label{eq:tcorr}
\end{align}\end{linenomath*}
where $I_S(\bm{q}, j, k)$ is the measured scattering intensity at reciprocal space coordinates $(q, \theta, \phi)$ and real-space coordinates $(j, k)$.
Here, $(j, k)$ are two Cartesian coordinates which give the position of the beam relative to the sample in the plane orthogonal to the incident beam direction, and $I_T$ is the transmitted intensity \cite{liebi_nat_2015}.
The location of the \gls{rsm} is given in the experimental system coordinates $(j, k, p)$, where $p$ is the coordinate of the direction in which the x-ray beam travels, see \autoref{fig:sphere_of_proj}a).

The subset of $\textsc{RSM}(\bm{q})$, which is possible to measure at a given sample orientation under the small-angle approximation, lies on a great circle given by
\begin{linenomath*}\begin{align}
    C(\varphi,\alpha,\beta) = \cos(\varphi)\bm{q}_0(\alpha, \beta) + \sin(\varphi)\bm{q}_{90}(\alpha, \beta)
    \label{eq:great_circle}
\end{align}\end{linenomath*}
where $\varphi$ is an angle on the detector, $\bm{q}_{0}(\alpha, \beta)$ and $\bm{q}_{90}(\alpha, \beta)$ are two unit vectors in the sample coordinate system aligned with the $0^\circ$ and $90^\circ$ direction of the detector, respectively, $(\alpha, \beta)$ are two angles which give the sample orientation as a sequence of rotations about two axes $\hat{\pmb{\alpha}}$ and $\hat{\pmb{\beta}}$ orthogonal to the direction of the impinging beam, see \autoref{fig:sphere_of_proj}a).

The sample is mounted on a rotation stage such that the first rotation $R_{\beta}$ also rotates $\hat{\pmb{\alpha}}$, and a sequential rotation of the sample may therefore be described by the composite rotation $R_\beta R_\alpha$.
We choose the sample coordinate system such that it coincides with the experimental coordinate system when $\alpha, \beta = 0$, see the initial sample mounting in \autoref{fig:sphere_of_proj}b).
To simplify this analysis without loss of generality, we will parameterize the measured reciprocal space vector as $\hat{\pmb{\alpha}} = \bm{q}_{90}(0, 0)$ and  $\hat{\pmb{\beta}} = \bm{q}_{0}(0, 0)$.
Then, the coordinate system of the sample will be subject to the composite rotation $R_{\bm{q_{0}}}(\beta)R_{\bm{q_{90}}}(\alpha)$, and the direction of the impinging beam in the sample coordinate system thus changes according to
\begin{linenomath*}\begin{align}
\bm{p}(\alpha, \beta) = R^T_{\bm{q_{90}}}(\alpha)R^T_{\bm{q_{0}}}(\beta)\bm{p}(0, 0),
\label{eq:vector_rotation}
\end{align}\end{linenomath*}
where $\bm{p}(0, 0)$ is the direction of the beam in the sample coordinate system prior to any rotation or tilt of the sample.
We can understand the values taken by $\bm{p}(\alpha, \beta)$ as points on a \textit{sphere of projection}, which is a unit sphere consisting of all unique measurement directions up to Friedel symmetry, see \autoref{fig:sphere_of_proj}d).
Equation \eqref{eq:vector_rotation} also applies to $\bm{q}_{90}$ and $\bm{q}_{0}$.

A composite rotation of the projection vector about the two axes $R_{\bm{q_{0}}}(\beta)R_{\bm{q_{90}}}(\alpha)$ may be described as a single rotation around a third axis $\hat{\imath}$, which lies on $C(\varphi, \alpha, \beta)$.
Consequently, this direction is a rotational invariant with respect to said composite rotation.
This means that $RSM(\Vert \bm{q} \Vert \hat{\imath}, \bm{r})$ can be regarded as a scalar quantity for the purpose of tomography, and standard tomographic analysis can therefore, in principle, be applied to the reconstruction of this component.
Although carrying out the experiment in practice requires a third rotation axis when remounting, as it is not possible to tilt the sample by more than $45^\circ$, it is possible to specify all points on the sphere of projection using rotations about only two orthogonal axes.
Any component of the rotation that occurs about the axis of projection can be discounted in this analysis, since it does not change the information contained in the projection.
The following line of reasoning is therefore also valid when combining data from the the two measurements.
According to the projection-slice theorem, the Fourier transform of a projection along $\bm{p}$ constitutes a slice orthogonal to $\bm{p}$ in Fourier space \citeaffixed{garces_pjt_2011}{\textit{e.g.}, }.
Tomographic reconstruction can therefore be understood as the problem of interpolating between slices in Fourier space.
This implies that a set of sufficiently densely placed projections along a great semicircle on the sphere of projection must be measured for a reconstruction of good quality of any given point on the \gls{rsm}.
This leads to the so-called missing wedge problem, where the absence of projections along any section of this great semicircle leads to a missing wedge in the Fourier transform of the reconstruction, and thus a blurring in that direction.

\begin{figure}
    \includegraphics[width=\linewidth]{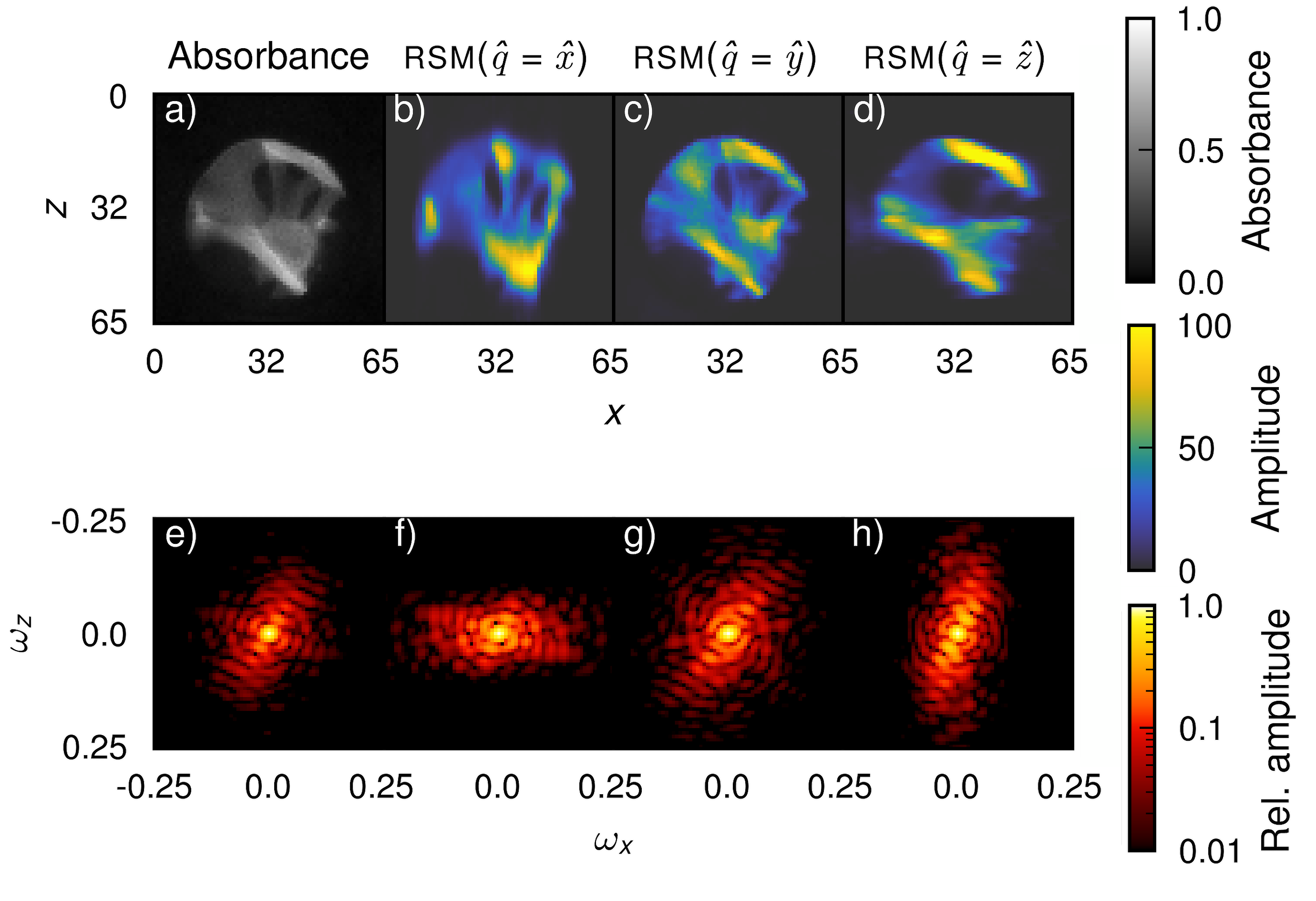}
  \caption{
    \textbf{Illustration of the missing wedge problem in \gls{saxstt}.}
    \textbf{a)--d)} show a projection along the $y$-direction of the component of reconstructions of, a), the absorbance, and, b)--d), the \gls{rsm} aligned with the the $x$-, $y$-, and $z$-directions, respectively.
    \textbf{e)--h)} show the amplitudes, normalized by the zero frequency component, of the discrete Fourier transforms of the projections as functions of the discrete frequency $\bm{\omega}$.
    Smearing of b) and d) occur in the $z$- and $x$-direction respectively, but no smearing can be seen for a) or c).
    }
  \label{fig:missing_wedges}
\end{figure}

In \autoref{fig:missing_wedges}, projections of reconstructions from data set 1 (\autoref{fig:sphere_of_proj}d)) along the $y$-direction of the absorbance as well as the reciprocal space map amplitude in three different directions are shown, along with the discrete Fourier transform of each projection.
The absorbance of each projection can be defined as
\begin{linenomath*}\begin{equation*}
    a(j, k) = -\log{\left(\frac{I_T(j, k)}{I_0(j, k)}\right)},
\end{equation*}\end{linenomath*}
where $I_T$ is the transmitted intensity at each point in the raster scan, and $I_0$ is the incident intensity, where we use the projection-wise approximation $I_0(j, k) \approx \max{I_T(j, k)}$, since each measurement includes some air, which has a very small absorbance.
The absorbance is a scalar quantity and therefore very accurately reconstructed without missing wedges.
Hence, it is included for comparison.
This projection direction, along the y-axis, is not part of the measurement of data set 1, as seen in  \autoref{fig:sphere_of_proj}b), which is why it is useful in illustrating missing wedges.
The x- and z-components of the \gls{rsm} shown in \autoref{fig:missing_wedges}b) and d) can be regarded as measurements which are missing from the data; the y-component in c) would not actually be measured along this projection direction but does not suffer from missing wedges, because it is measured from every orthogonal direction (the measurements which lie on the line where $y = 0$ in \autoref{fig:sphere_of_proj}d)).
We can observe how the amplitudes in \autoref{fig:missing_wedges}b) and d) are smeared out in the directions orthogonal to the reciprocal space map component compared to a) and c).
Equivalently, the discrete Fourier transforms are attenuated in the directions of smearing, relative to the more symmetric Fourier transforms in e) and g).
The attenuated segments of the discrete Fourier transforms correspond to missing orthogonal projections, per the projection-slice theorem.

A set of reconstruction constraints similar to those given by the projection-slice theorem exist for the general case of three-dimensional projections in the form of John's equation \citeaffixed{ma_je_2017}{\textit{e.g.}, }, which has been generalized to the case of arbitrary-rank symmetric tensor fields, resulting in additional smoothness constraints on the components of the tensor field \cite{sharafutdinov2012integral,nadirashvili_2016_iop}.
Therefore, to treat this problem more generally, and not just for discrete, precisely measured \gls{rsm} components, we need to assume that the reciprocal space map does not change too quickly across real- and reciprocal space dimensions.
Then, given the existence of the invariant axis $\hat{\imath}$, we can define a sampling quality factor on the reciprocal space sphere based on the density of sampling on the sphere of projection.
To accomplish this, we define a Friedel symmetric sampling density $\rho(\bm{p}(\alpha, \beta))$ as
\begin{linenomath*}\begin{align*}
    \rho(\bm{p}(\alpha, \beta)) = \begin{cases}
    1&\text{ if the direction $\bm{p}(\alpha', \beta')$ of the nearest}\\
    \ &\text{measurement satisfies $\Delta(\bm{p}(\alpha, \beta), \bm{p}(\alpha', \beta')) < \delta$},\\
    0&\text{ otherwise.}
    \end{cases}
\end{align*}\end{linenomath*}
with $\bm{p}(\alpha, \beta)$ being defined by Eq.~\eqref{eq:vector_rotation}, and where $\Delta(\bm{v}, \bm{u})$ is the Friedel symmetric great-circle distance, defined as
\begin{linenomath*}\begin{equation}
    \Delta(\bm{v}, \bm{u}) = \arccos\left({\frac{\vert \bm{v} \cdot \bm{u}\vert }{\Vert \bm{v} \Vert \Vert \bm{u} \Vert}}\right),
    \label{eq:great_circle_distance}
\end{equation}\end{linenomath*}
where $\bm{v}$ and $\bm{u}$ are two vectors.
The precise choice of the threshold parameter $\delta$ depends on assumptions about both the real and reciprocal space continuity of the sample, the size of the sample, as well as the chosen reconstruction method.
We chose $\delta$ under the assumption that our sampling density in well-sampled regions was sufficient to obtain a good tomographic reconstruction.
This was chosen over more quantitative thresholds such as the great-circle distance implied by the Nyquist-Shannon sampling theorem applied to tomographic reconstruction \cite{nw_math_2001} for several reasons.
First, we carry out the reconstruction under smoothness and sparsity constraints, which reduce the required number of samples.
Second, John's equation for tensor tomography complicates the assumptions based on which standard sampling factors are calculated, since they will also depend on smoothness in reciprocal space \cite{nadirashvili_2016_iop}, and consequently using standard measures would give a misleading impression of certainty about the required number of samples.
Therefore, to avoid over-complicating the analysis, we prefer to use $\rho(\bm{p}(\alpha, \beta))$ as a relative measure of sampling density.

\begin{figure}
\includegraphics[width=0.5\linewidth]{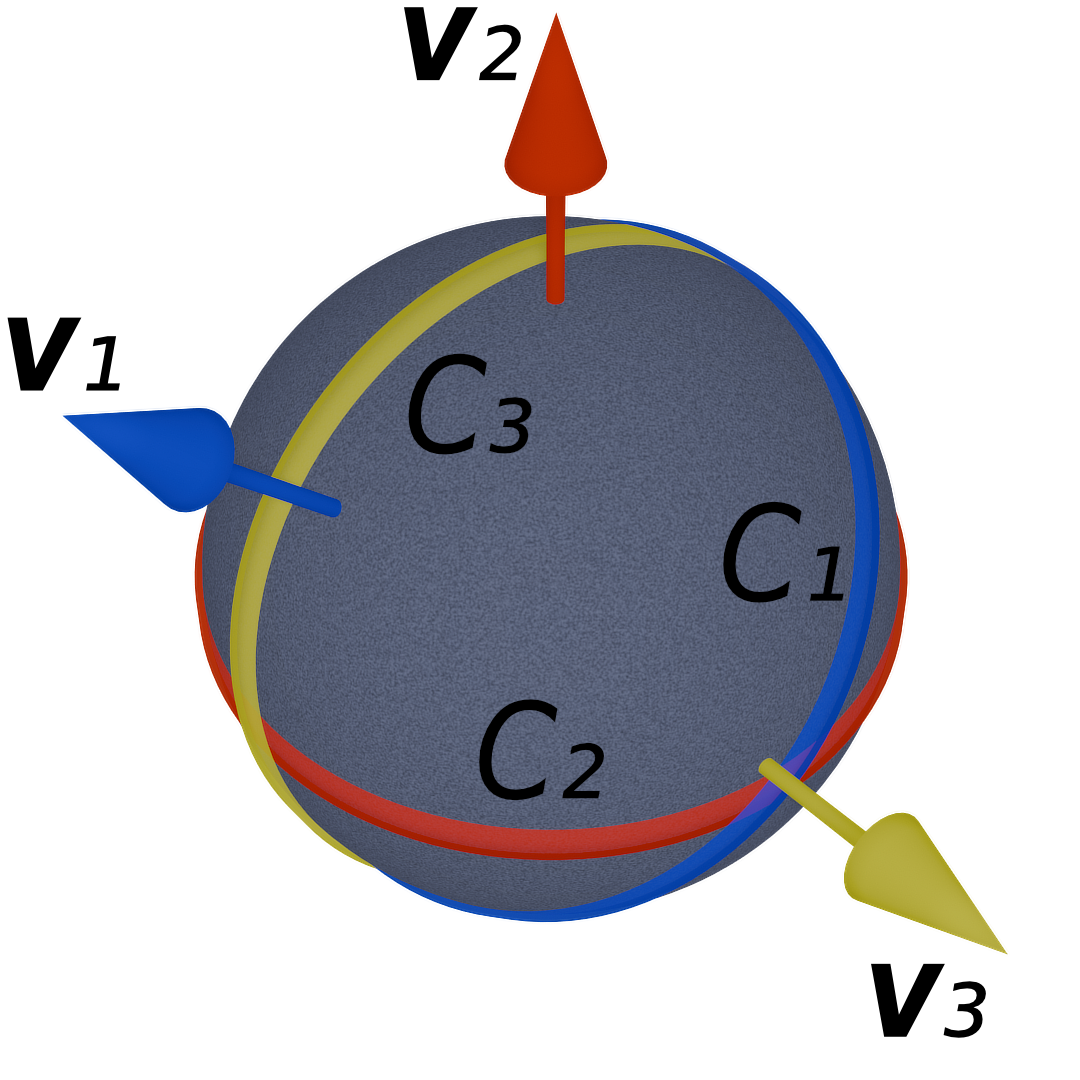}
  \caption{
    \textbf{Examples of vectors defining directions on a sphere, and their respective orthogonal great circles.}
    The three vectors labelled $\bm{v}_1$ (blue), $\bm{v}_2$ (red), and $\bm{v}_3$ (yellow) each have a unique orthogonal great circle $C_1$, $C_2$, and $C_3$.
    }
  \label{fig:sphere_gc}
\end{figure}

Because we sample points on the reciprocal space sphere orthogonal to the respective points on the sphere of projection, we can compute a quality factor by evaluating the Funk-Radon transform, i.e. the normalized integral over the orthogonal great circle \cite{Funk_1913}, of the sampling density $\rho$.
In other words, the quality factor can be computed as
\begin{linenomath*}\begin{align}
    F[\rho](\theta, \phi) = \frac{1}{2\pi}\int_0^{2\pi} \mathop{d\tau}\rho(C(\tau,\theta,\phi)),
    \label{eq:funkradon}
\end{align}\end{linenomath*}
where $(\theta, \phi)$ specifies a direction in reciprocal space, and $C(\tau, \theta, \phi)$ is defined as in Eq.~\eqref{eq:great_circle}.
Although we defined $C(\tau, \theta, \phi)$ as giving the directions in reciprocal-space measured, given a real-space direction, the symmetry of this relationship means that we can invert it to give real-space directions to measure, given a reciprocal-space direction that we wish to probe.
The relationship between directions on a unit sphere and their unique orthogonal great circle is shown in \autoref{fig:sphere_gc}.
Applying Eq.~\autoref{eq:funkradon}, the value of the quality factor at, \emph{e.g.}, $\bm{v}_1$ would be defined by the integral of $\rho$ over all directions in $C_1$.
Using this quality factor, which lies in the range $[0, 1]$, where $0$ means the lowest possible quality and $1$ means the highest possible quality, we may predict the quality of the tomographic reconstruction at any point in reciprocal space.

In order to obtain a reconstruction suitable for the analysis of the missing wedge problem, we want to represent the \gls{rsm} on the sphere using smooth local functions on the sphere which can be projected into our measurement basis.
The locality of the representation is important, since non-local artefacts (such as the spherical harmonic Gibbs phenomenon) could otherwise affect the evaluation of the reconstructions in unpredictable fashion.
We also want to reduce the measured data into azimuthal bins in order to keep the data size manageable, and this reduction can be represented by integrating $I_S(\left\Vert{\bm{q}}\right\Vert C(\varphi, \alpha, \beta), j, k)$ (Eq. \eqref{eq:tcorr}) over segments of $\varphi$ \citeaffixed{bunk2009multimodal}{as in, e.g., }.
\citeasnoun{schaff_nature_2015} utilized the existence of an invariant point in the \gls{rsm} for any given rotation, referring to this point as a ``virtual axis''. 
However, \citeasnoun{schaff_nature_2015} employed only limited reduction of the measured \gls{rsm}, necessitating extensive sorting of measurements according to their nearest virtual axis, and processing of a large number of separate tomographic problems, followed by subsequent composition and analysis of the separate reconstructions.
\citeasnoun{defalco_2021_jac} also utilized rotational axis invariance, and carried out a reconstruction using the component of the \gls{saxs} measurement orthogonal to the main axis of rotation to study a subpopulation of mineral particles within a sample.
These approaches utilize the separability of measurements in order to simplify the tomographic reconstruction problem.
However, azimuthal binning reduces this separability, unless the bins are made very small, which would work against the purpose of reducing data size.
Thus, rather than aiming to carry out reconstructions at specific points in reciprocal space and subsequently fitting a function to these points, we define a grid of Gaussian basis functions on the unit sphere.
This basis set forms a local representation of spherical functions which is used to interpolate measured data into a smooth function on the sphere \cite{fornberg_spherical}.
Gaussian radial basis functions have an advantage over the spherical harmonic representation used in \citeasnoun{nielsen_tt_2023} because they do not suffer from the Gibbs phenomenon or other non-local artefacts \cite{gelb_1997_gibbs}.
We define a set of projection matrices from the spherical \gls{rsm} to the detector by left-multiplication as
\begin{linenomath*}\begin{align}
    G_{i,nm} &= \frac{1}{\mathcal{N}_{n} \left\vert\varphi_{m+1} - \varphi_{m}\right\vert}\int\displaylimits^{\varphi_{m+1}}_{\varphi_{m}} \mathop{d\tau}\exp\left(\frac{\Delta((1, \theta_n, \phi_n), C(\tau, \alpha_{i}, \beta_{i}))^{2}}{2\sigma^2}\right),
    \label{eq:local_repr}
\end{align}\end{linenomath*}
where $\mathcal{N}_{n}$ is a normalization factor, $\left[\varphi_m, \varphi_{m + 1}\right)$ parameterizes the $m$th detector segment on the unit circle, $(\alpha_i, \beta_i)$ gives the sample orientation, $(1, \theta_n, \phi_n)$ is a unit vector expressed in spherical coordinates giving the location of the mode of basis function $n$, $\sigma$ parameterizes the width of each basis function, $\Delta(\bm{u}, \bm{v})$ for any two vectors $(\bm{u}, \bm{v})$ is the great-circle distance defined by Eq.~\eqref{eq:great_circle_distance}, and finally $C(\tau, \alpha_i, \beta_i)$ is defined by Eq.~\eqref{eq:great_circle}, with $\tau$ being an integration variable that parameterizes the integration over each segment.
The normalization factor $\mathcal{N}_{n}$, which evens out irregularities in the distribution of grid points, is given by the sum of all rows in an auto-projection matrix $G_{nn'}$, which can be expressed in a similar form as \autoref{eq:local_repr} but evaluated only at one point rather than integrated over,
\begin{linenomath*}\begin{align*}
    \mathcal{N}_n = \sum_{n'}G_{nn'} &= \sum_{n'}\exp\left(\frac{\Delta\left((1,\theta_n, \phi_n), (1,\theta_{n'}, \phi_{n'})\right)^{2}}{2\sigma^2}\right),
\end{align*}\end{linenomath*}
where $n$ and $n'$ both run the indices of all basis functions.
In this work, the basis functions have been distributed on a modified Kurihara mesh \cite{kurihara_mwr_1965}, with an approximately equal distribution over the unit sphere.
The modified Kurihara mesh depends on an integer scale parameter $s$ which determines the number of basis functions on the hemisphere, according to $N = 2s^2$.
The width parameter was chosen based on a simple heuristic for smooth and non-oscillatory interpolation, $\sigma = \frac{\pi}{2 \cdot s}$.
We chose $s = 9$ as the scale parameter, thus yielding $N = 162$ basis functions and width parameter $\sigma = \frac{\pi}{18}$.
These choices yield smoothly interpolated functions without oscillations, and a density of basis functions greater than the density of detector segments, but smaller than the density of projection directions.
For more details on the modified Kurihara mesh and the basis functions, see \ref{sect:gaussian_kernels}.
Since Gaussians do not have compact support, i.e., they do not fall off to zero, the kernels are local only in a non-strict sense -- the vast majority of the amplitude of each basis function is located within a small area around its mode, assuming the standard deviation $\sigma$ is at least a few times smaller than $\frac{\pi}{2}$.
Because of this locality property, we expect the reliability of the coefficient of the basis function located at $(\theta_n, \phi_n)$, considered across real space, to be related to the value of the quality factor $F[\rho](\theta_n, \phi_n)$ given by Eq.~\eqref{eq:funkradon}.

Completing the description of the forward model requires the definition of a John transform matrix for tensor tomography, which is treated in greater detail in \citeasnoun{nielsen_tt_2023}.
The resulting set of matrices $\mathbf{P}_i$ together define a transform of a tensor field in three-dimensional space into a tensor field in projection space, with each $i$ indicating a projection direction, similarly to how $\mathbf{G}_i$ (Eq.~\eqref{eq:local_repr}) describes a transform between detector space and spherical space.
This allows us to describe the system of equations to be solved for each projection $i$ as
\begin{linenomath*}\begin{align}
	\mathbf{P}_i\bm{X}\mathbf{G}_i = \bm{D}_i,
    \label{eq:system}
\end{align}\end{linenomath*}
where $\bm{D}^i$ is matrix of data measured from a single projection.

\section{Methods}
\label{sect:methods}
\subsection{Formalism}
\label{sect:formalism}
In order to improve the rate of convergence of the system in Eq.~\eqref{eq:system}, we compute a series of weight and preconditioning matrices.
Each weight matrix is computed as
\begin{linenomath*}\begin{align*}
    \mathbf{W}_i = (\mathbf{P}_i\bm{U}\mathbf{G}_i)^{\circ (-1)},
\end{align*}\end{linenomath*}
where $U$ is a matrix filled with the value $1$ everywhere and $\mathbf{A}^{\circ (-1)}$ denotes a relaxed element-wise multiplicative inverse of $\mathbf{A}$,
\begin{linenomath*}\begin{align*}
    [\mathbf{A}^{\circ (-1)}]_{ij} = 
        \begin{cases}
        A_{ij}^{-1} \text{, if } A_{ij} \geq \epsilon,\\
        \epsilon^{-1} \text{ otherwise.}
    \end{cases}
\end{align*}\end{linenomath*}
for some predefined $\epsilon > 0$. Similarly, each preconditioning matrix is computed as
\begin{linenomath*}\begin{align*}
    \mathbf{C}_i = (\mathbf{P}^T_i\bm{V}\mathbf{G}^T_i)^{\circ (-1)},
\end{align*}\end{linenomath*}
where $V$ is a matrix filled with the value $1$ everywhere.
We may now write the system to be solved for each projection $i$ as
\begin{linenomath*}\begin{align*}
    \mathbf{C}_i \odot (\mathbf{P}^T_i (\mathbf{W}_i \odot  (\mathbf{P}_i\bm{X}\mathbf{G}_i)) \mathbf{G}_i^T) =  \mathbf{C}_i \odot (\mathbf{P}^T_i (\mathbf{W}_i \odot \bm{D}_i) \mathbf{G}_i^T)
\end{align*}\end{linenomath*}
where $\odot$ denotes the Hadamard or elementwise product.
This is analogous to the weights and preconditioner used in the SIRT algorithm for scalar tomography \citeaffixed{gregor_2015_ieee}{see \emph{e.g.}, }, which was utilized by \citeasnoun{schaff_nature_2015} in the separate reconstructions about each virtual axis.
This weight- and preconditioner pair serves to normalize the gradient by accounting for the number of voxels that contribute to each pixel, and the number of projections that contribute to each voxel.
This normalization is done for each detector segment and each \gls{rsm} basis function, accounting also for the detector-to-sphere mapping of Eq.~\eqref{eq:local_repr}.
This system is then solved through least-squares Nestorov-accelerated gradient descent, subject to coefficient-wise total variation and $L_1$ norm regularization, optimized in the Huber approximation of each \cite{huber_1964}, using existing implementations in the \textsc{mumott} package \cite{nielsen_tt_2023,nielsen_2024_10708583}.
The reconstruction of the full data set took \qty{630}{\second}, and the reconstruction of each partial data set took \qty{380}{\second}, on a workstation using an Nvidia RTX 3060 GPU, an 8-core AMD Ryzen 7 3700X CPU, and 64 GB DDR4 2666 MHz RAM.

\subsection{Computations}
The integral in Eq.~\eqref{eq:local_repr} was computed by quadrature utilizing the adaptive Simpson's rule \citeaffixed{lyness_simpson}{\emph{e.g.}, }, terminating when the largest change in a matrix element, relative to the largest element in the matrix, fell below \num{e-5}.
Analysis of the orientation (\autoref{fig:ori_error}), the Funk-Radon Transform (Eq.~\eqref{eq:funkradon}), and computation of the scalar quantities in \autoref{fig:scalars} requires transforming the spherical function representation from a local Gaussian kernel representation to a spherical harmonic representation.
This is done by Driscoll-Healy quadrature \cite{driscoll_healy_1994}, sampling the function by evaluating the representation on a dense curvilinear grid.
The mean amplitude, and the relative anisotropy, are defined as in \citeasnoun{nielsen_tt_2023}, i.e., as the spherical mean and the spherical standard deviation normalized by the mean.
These figures of merit are similar to the ``symmetric intensity'' and ``degree of orientation'' used by, e.g., \citeasnoun{bunk2009multimodal}.
The fiber symmetry factor is given by
\begin{linenomath*}\begin{align}
    S(\bm{a}) = \frac{\sqrt{\sum_{\ell=1}\sum_{m=-\ell}^\ell F[\bm{a}]_\ell^m\hat{Y}_\ell^m(\theta, \phi)}}{\sqrt{\sum_{\ell=1}\sum_{m=-\ell}^\ell F[\bm{a}]_\ell^m} \sqrt{\sum_{\ell=1}\sum_{m=-\ell}^\ell \hat{Y}_\ell^m(\theta, \phi)}},
    \label{eq:fibersym}
\end{align}\end{linenomath*}
where $\bm{a}$ is the spherical harmonic representation of a \gls{rsm}, $F[\cdot]$ is the Funk-Radon transform, $\hat{\bm{Y}}_\ell^m$ is a spherical harmonic basis function, and $(\theta, \phi)$ is the orientation of the \gls{rsm}, as given by the minimal eigenvector of its rank-2 tensor representation.
This figure of merit evaluates how similar the \gls{rsm} is to an ideal ring function (which has all of its amplitude at a great circle consisting of the points orthogonal to its orientation).
Consequently, it quantifies the extent to which a \gls{rsm} exhibits the equatorial symmetry expected from diffuse mineral scattering in bone.

The orientation error is computed as $\Delta(\bm{v}, \bm{u})$ (Eq.~{\eqref{eq:great_circle_distance}}), where $\bm{v}$ and $\bm{u}$ are two orientation vectors.
The orientation error is thus the angle subtended by the two orientation vectors, accounting for the Friedel symmetry of orientation vectors.
\subsection{Implementation}
The version of \textsc{mumott} used in this work can be found at the DOI \href{https://doi.org/10.5281/zenodo.10708583}{10.5281/zenodo.10708583} \cite{nielsen_2024_10708583}.
New versions of \textsc{mumott} are continuously made available at the DOI \href{https://zenodo.org/doi/10.5281/zenodo.7919448}{10.5281/zenodo.7919448}.
The John transform in \textsc{mumott} is implemented using a bilinear interpolation algorithm which supports multiple channels per voxel and pixel, written using the CUDA API of the Python package Numba \cite{lam_2015_numba}.
The algorithm employed is based on the work of \citeasnoun{xu_jt_jsb} and \citeasnoun{palenstijn_jt_jsb}.
Other computations were carried out using the Python packages NumPy and SciPy \cite{numpy,scipy}.
Two-dimensional plots were created using the package Matplotlib \cite{hunter_2007}.
The color maps used in this work are from the package ColorCET \cite{kovesi_2015, kovesi_2020}.
The experimental setup render in \autoref{fig:sphere_of_proj} was created using Blender \cite{blender_2018}. 
All other 3D renders in this work were created using ParaView \cite{paraview}.
The two data sets were aligned using the cross-correlation algorithm of \citeasnoun{manuel_subpixel}, and rotated by modifying the set of vectors used to define the reconstruction geometry in \textsc{mumott}.
The rotations were first determined by eye and then refined by comparing absorptivity reconstructions in ParaView.

\section{Experiment}
\label{sect:Experiment}
The sample chosen for this study was trabecular bone fixed and embedded in \gls{pmma}.
A cube was extracted from the bulk and subsequently milled into a cylinder of diameter \qty{1.2}{\milli\meter} and height \qty{1.2}{\milli\meter} using a custom-made lathe system \cite{holler_jsr_lathe}.
The sample was measured at the cSAXS beamline of the Swiss Light Source (SLS) at the Paul Scherrer Institut (PSI), Switzerland. The X-ray energy was set to \qty{12.4}{\kilo\eV} using a \ce{Si} \hkl(1 1 1) double crystal monochromator, and the scattering patterns were recorded on a Pilatus 2M detector placed at a sample to detector distance of  \qty{2.17}{\meter}.
A flight tube, approximately \qty{2}{\meter} in length, was placed in between the sample and detector to place to reduce the air scattering.
A \qty{1.5}{\milli\meter} steel beamstop inside the flight tube blocked the directly transmitted beam.
The fluorescence signal from the beamstop, proportional to the intensity of the impinging x-rays ($I_T$ in Eq.~\eqref{eq:reciprocal_space_map}), was measured by a Cyberstar (Oxford Danfysik).
This allowed the relative x-ray transmission through the sample to be measured.
The sample was measured with a beam that had full-width half maxima of \qty[parse-numbers=false]{12 \times 24}{\micro\meter\squared} as measured by a knife-edge scan.
The raster scan used a step size of \qty{25}{\micro\meter} in both the vertical and horizontal directions, with continuous fly-scanning in the vertical direction.
The experimental setup is illustrated in \autoref{fig:sphere_of_proj} a).
Two sets of \gls{saxstt} measurements were carried out, each consisting of \num{224} scanning \gls{saxs} images. 
The two sets of measurements are shown on the sphere of projection in \autoref{fig:sphere_of_proj} d) and e), where each marker indicates the direction of the x-ray beam (given by $\bm{p}(\alpha, \beta)$ in Eq.~\eqref{eq:vector_rotation}) in the sample coordinate system.

During the first set of \gls{saxstt} measurements, the base of the cylinder sample was glued to the end of a \gls{pmma} needle, using a hot water-soluble glue (Norland Blocking Adhesive 107).
Before the second \gls{saxstt} experiment, the sample was glued with UV-glue (Norland Optical Adhesive 81) to a second pin, before detaching the first pin by placing the sample in hot water.
The second pin was placed at approximately \qty{90}{\degree} to the first pin, measured around the axis of the initial direction of the beam, see \autoref{fig:sphere_of_proj}.
In total, \num{1716960} scattering images were measured.

\begin{figure}
    \includegraphics[width=\linewidth]{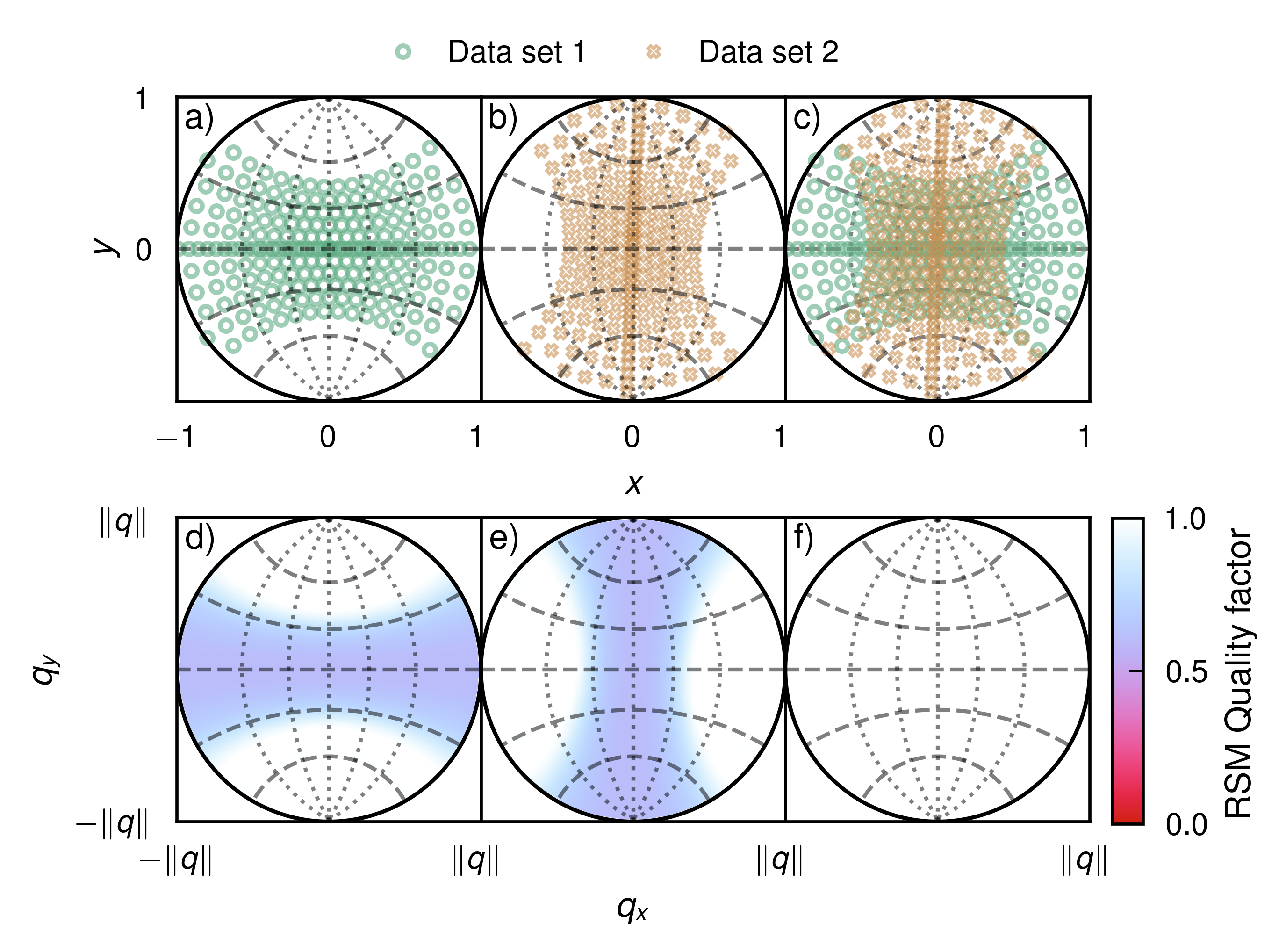}
  \caption{
    \textbf{Points on hemisphere of projection and theoretical quality factors.}
    \textbf{a)} Probed points on sphere of projection in first measurement.
    \textbf{b)} Probed points in second measurement.
    \textbf{c)} Combined points from both measurements.
    \textbf{d)} Quality factor in reciprocal space from first data set.
    \textbf{e)} Quality factor from second data set.
    \textbf{f)} Quality factor from combined data sets.
    The dotted lines show great circles at longitudes $0^\circ$, $\pm 30^\circ$ and $\pm 60^\circ$ with the $y$-axis as the meridian.
    The dashed lines show small circles with elevations of $0^\circ$, $\pm 30^\circ$ and $\pm 60^\circ$ with the $x$-axis as the equator.
    }
  \label{fig:samples}
\end{figure}

\autoref{fig:samples}a--c) shows the directions of measurement on the unit hemisphere of projection while \autoref{fig:samples}d--f) shows the quality factor $F$ defined by Eq.~\eqref{eq:funkradon} on the reciprocal space hemisphere.
Note that Friedel symmetry is accounted for in the hemispheric representation.
The reciprocal space quality factors follow the expected symmetry, where measurements along the entirety of a great semicircle result in a quality factor of \num{1} at the point orthogonal to this semicircle.
The lowest obtained quality factor is \num{0.5}, since the lowest possible coverage (for data set 1) of a great semicircle occurs when the semicircle lies at a single longitude and varies only in latitude.
Such a semicircle is still covered by measurements at a fixed longitude, the latitude (tilt, for data set 1) of which span the range [\qtyrange[range-phrase=\ensuremath{,{}},range-units=repeat]{-45}{45}{\degree}] --- thus, in the worst-case scenario, half of the semicircle's total arc length of $180^\circ$ is covered.

For the reconstruction and analysis, a $q$-range of \qtyrange{0.597}{0.607}{\per\nano\meter} was used, corresponding to a $d$-spacing range of \qtyrange{10.36}{10.53}{\nano\meter}.
This range was used due to artefacts in the second measurement at lower $q$-ranges, possibly due to the water-soluble glue penetrating the outer layer during the remounting of the sample; see \ref{sect:supp_artefacts}, as well as \ref{fig:raster_scans}, and \ref{fig:volrend_diff_sym} for details.

\section{Results}
\label{sect:results}

\begin{figure}
    \includegraphics[width=\linewidth]{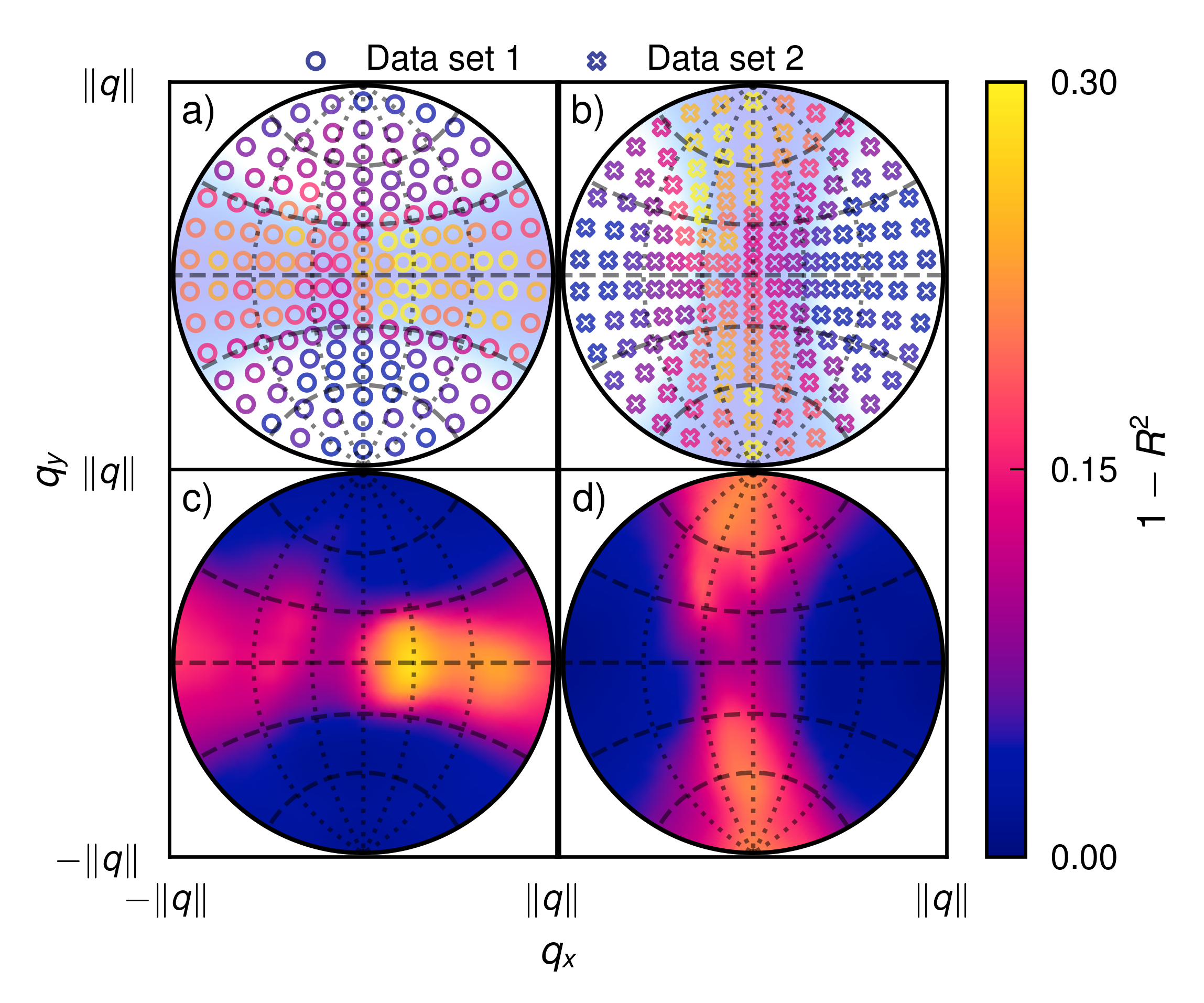}
  \caption{
    \textbf{Error distribution for each \gls{rsm} basis function.}
    \textbf{a)} Coefficient errors for data set 1.
    \textbf{b)} Coefficient errors for data set 2.
    \textbf{c)} \Gls{rsm} errors for data set 1.
    \textbf{d)} \Gls{rsm} errors for data set 2.
    The markers have been placed over the corresponding theoretical quality factor from \autoref{fig:samples}.
    The errors in \textbf{a)} and \textbf{b)} were calculated by computing the overall Pearson correlation coefficient for each basis function coefficient when comparing the partial and full data sets.
    In \textbf{c)} and \textbf{d)} the correlation coefficients were computed for the \gls{rsm} amplitude at each coordinate.
    The correlation factors for the coefficients and the \gls{rsm} amplitudes are not the same, since the Gaussian radial basis functions of the basis set overlap.
    The dotted lines show great circles at longitudes $0^\circ$, $\pm 30^\circ$ and $\pm 60^\circ$ with the $y$-axis as the meridian.
    The dashed lines show small circles with elevations of $0^\circ$, $\pm 30^\circ$ and $\pm 60^\circ$ with the $x$-axis as the equator.
  }
  \label{fig:error}
\end{figure}

The results of comparing a reconstruction of the full data set, which combines data sets 1 and 2, with reconstructions that include, respectively, only data set 1 or 2, are shown in \autoref{fig:error}.
The location of each marker in a) and b) corresponds to the mode of a \gls{rsm} basis function, see Eq.~\eqref{eq:local_repr}, while the color of the marker corresponds to the error computed from comparing the coefficients of that basis function to the corresponding coefficients of the full dataset reconstruction.
The markers are overlaid over the reciprocal space quality factor.
In c) and d), the error for the amplitude at each point on the reciprocal space sphere is shown.
The distribution of errors in the reciprocal space amplitude follows the quality factor closely, with errors above approximately \num{0.1} occurring exclusively in the region where the quality factor is smaller than \num{1}.
The errors for the basis set coefficients in a) and b) are larger than the errors of the amplitude in c) and d), which is especially apparent when comparing the upper region of a) to the same region in c).
This is explained by the fact that the basis set functions are not orthogonal but overlap.
This means that some variations in the basis set coefficients cancel out when the amplitude of each \gls{rsm} function is evaluated for the calculation of the error.

\begin{figure}
    \includegraphics[width=\linewidth]{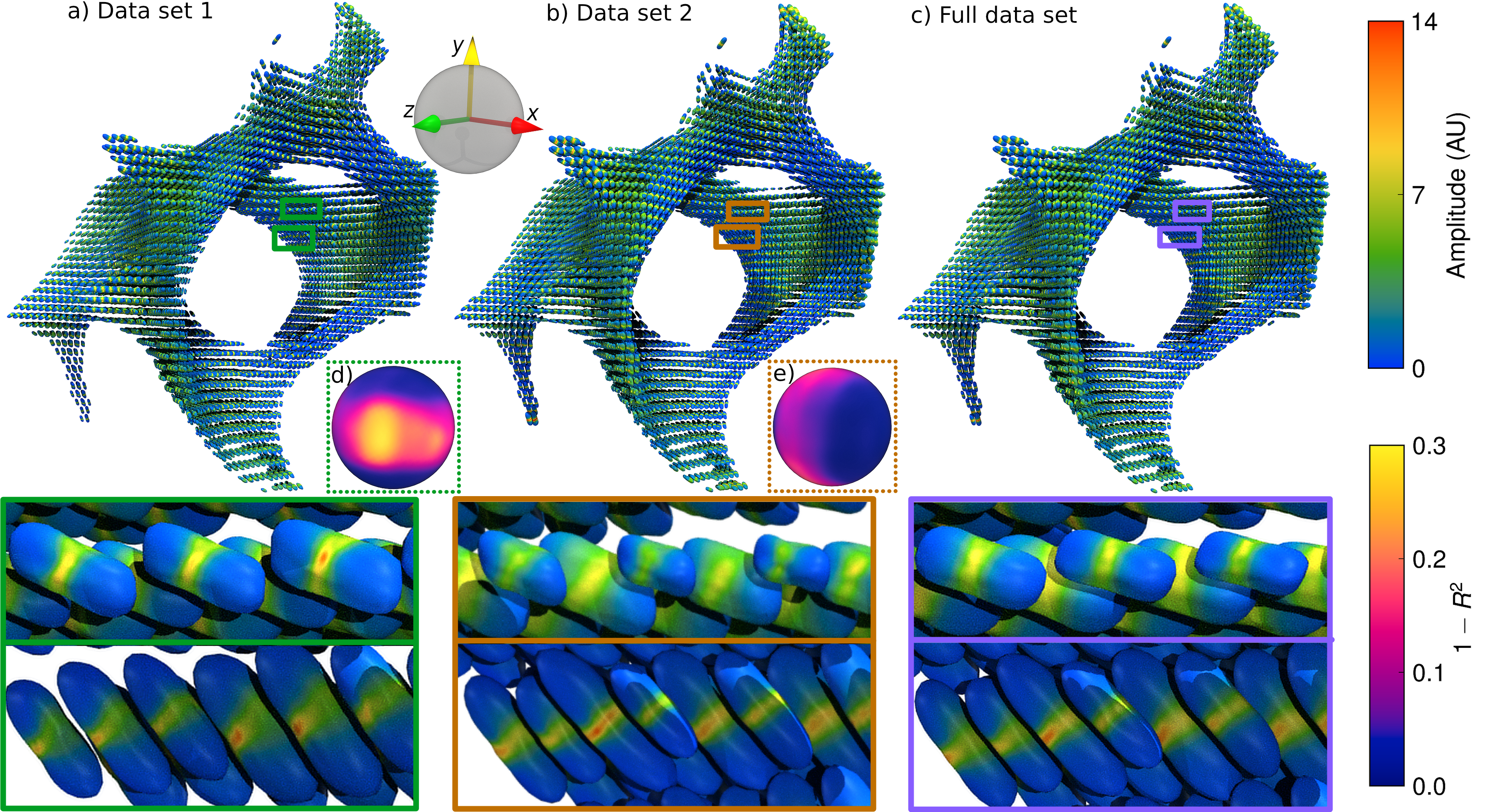}
  \caption{
    \textbf{Spherical function glyph render of reconstructions.}
    \textbf{a)} Partial data set 1, and a reciprocal space sphere showing the error distribution compared to the full data set.
    \textbf{b)} Partial data set 2, and a reciprocal space sphere showing the error distribution compared to the full data set.
    \textbf{c)} Full data set.
    \textbf{d)} Error distribution in reciprocal space for data set 1.
    \textbf{e)} Error distribution in reciprocal space for data set 2.
    The color of each spherical function indicates the \gls{rsm} amplitude, whereas the shape indicates the orientation.
    The shape is computed from the Funk-Radon transform of the \gls{rsm} amplitude.
    The insets show two sets of \glspl{rsm} that the partial reconstructions each have difficulty reconstructing, compared to the full data.
    }
  \label{fig:recon_tensors}
\end{figure}

In \autoref{fig:recon_tensors} the reconstructed \glspl{rsm} can be seen in a spherical function glyph render for a) data set 1 only, b) data set 2 only, and c) the full data set.
Each rendered glyph shows the \gls{rsm} reconstructed in that voxel, colored by its amplitude, and scaled by the Funk-Radon transform of the amplitude.
Because the scattering at this $q$-range is dominated by diffuse equatorial mineral scattering, deforming each glyph by the Funk-Radon transform allows its shape to visually indicate the orientation of the underlying nanostructure.

As illustrated in the insets d) and e), which show the error distribution on the \gls{rsm}, data set 1 has the best sampling and therefore the most reliable reconstruction along the $y$-axis, i.e., along the main tomographic axis (\autoref{fig:sphere_of_proj}).
Data set 2 has the smallest error along the $x$-axis.
The effect of this on the reconstructed 3D \gls{rsm} is illustrated in the enlarged views below each render.
The respective upper enlarged views shows \glspl{rsm} that are better reconstructed by data set 1, as data set 2 has difficulty reconstructing the amplitude near the $y$-axis, leading to increased asymmetry in the equatorial scattering due to missing wedges.
The lower enlarged views show \glspl{rsm} which are better reconstructed by data set 2, as data set 1 has difficulty reconstructing amplitudes near the $x$-axis, introducing additional texture in the equatorial scattering.
Both data sets have some difficulty reconstructing amplitudes that lie along the $z$-axis, but the difficulty is overall greater for data set 1, as indicated by the distribution on the spherical inset d), compared to the spherical inset e), which shows the amplitude error from \autoref{fig:error}, panels c) and d), respectively, rendered on a spherical surface.

It is likely that the primary explanation for this larger error is that the measurements close to the $x$-axis on the sphere of projection which result in the large errors near the $z$-axis must pass through the thickest part of the sample.
This means that the transmission is small, around \qty{4}{\percent}, compared to values of \qtyrange[range-units=repeat]{20}{50}{\percent} for thinner parts of the sample.
Consequently, noise in the transmission will have a relatively large impact on these measurements, see Eq.~\eqref{eq:tcorr}.

\begin{figure}
    \includegraphics[width=\linewidth]{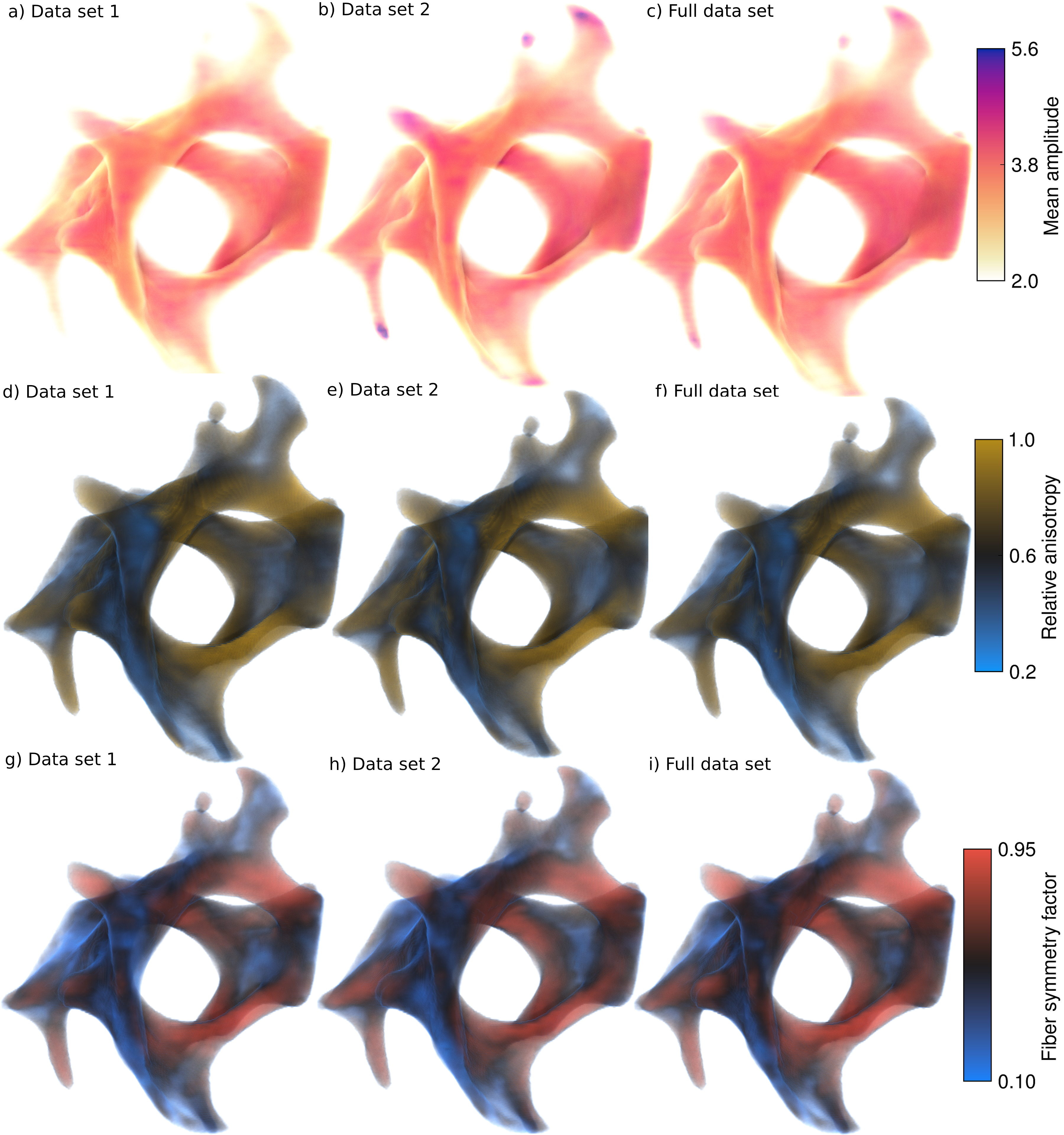}
  \caption{
    \textbf{Volume renders of scalar quantities.}
    \textbf{a) -- c)} Mean amplitude of \glspl{rsm} for partial data sets 1 and 2, as well as for the full data set.
    \textbf{d) -- f)} Relative anisotropy of \glspl{rsm}.
    \textbf{g) -- i)} Fiber symmetry factor of \glspl{rsm}.
  }
  \label{fig:scalars}
\end{figure}
Three scalar quantities for each of the three reconstructions can be seen in \autoref{fig:scalars}: the mean \gls{rsm} amplitude, the relative anisotropy (similar to quantities often referred to as \textit{degree of orientation}), and a fiber symmetry factor.
The fiber symmetry factor quantifies the degree to which the scattering is equatorial, see \nameref{sect:methods} and Eq.~\eqref{eq:fibersym} for details.
These quantities are of interest in evaluating the \glspl{rsm}, and therefore their similarity between partial and full data set reconstructions are of importance in evaluating the impact of the missing wedge problem.
The mean amplitude in the top row shows no large variations, except for slightly higher values at the edges of protrusions in data set 2, which may be due to the leeching of water-soluble glue into the sample during remounting, see \ref{sect:supp_artefacts}.
The mean amplitude is an important scalar value which is used for $q$-resolved reconstruction and further analysis of nanostructure information contained in the \gls{saxs} curve \cite{liebi_biomaterialia_2021, casanova_2023_biomaterials, silvabaretto_acta_2024}.
The relative anisotropy is also very similar for all three reconstructions, with almost no discernible differences.
Somewhat greater differences can be seen in the fiber symmetry factor, especially in the right-hand-side interface region where the insets in \autoref{fig:recon_tensors} are located.
The full data set has a high fiber symmetry factor in this area except at the very center of this interface, whereas the partial data sets appear to have a lower factor around the edges.
Thus, the fiber symmetry factor is more sensitive to missing wedges than the ordinary relative anisotropy.
This can also be seen in \autoref{fig:recon_tensors}, where additional texture within the ring of the equatorial scattering appears as an artefact of the missing wedge.
For quantitative plots of the distribution of the quantities, see \ref{sect:supp_abs_dev}.

\begin{figure}
    \includegraphics[width=\linewidth]{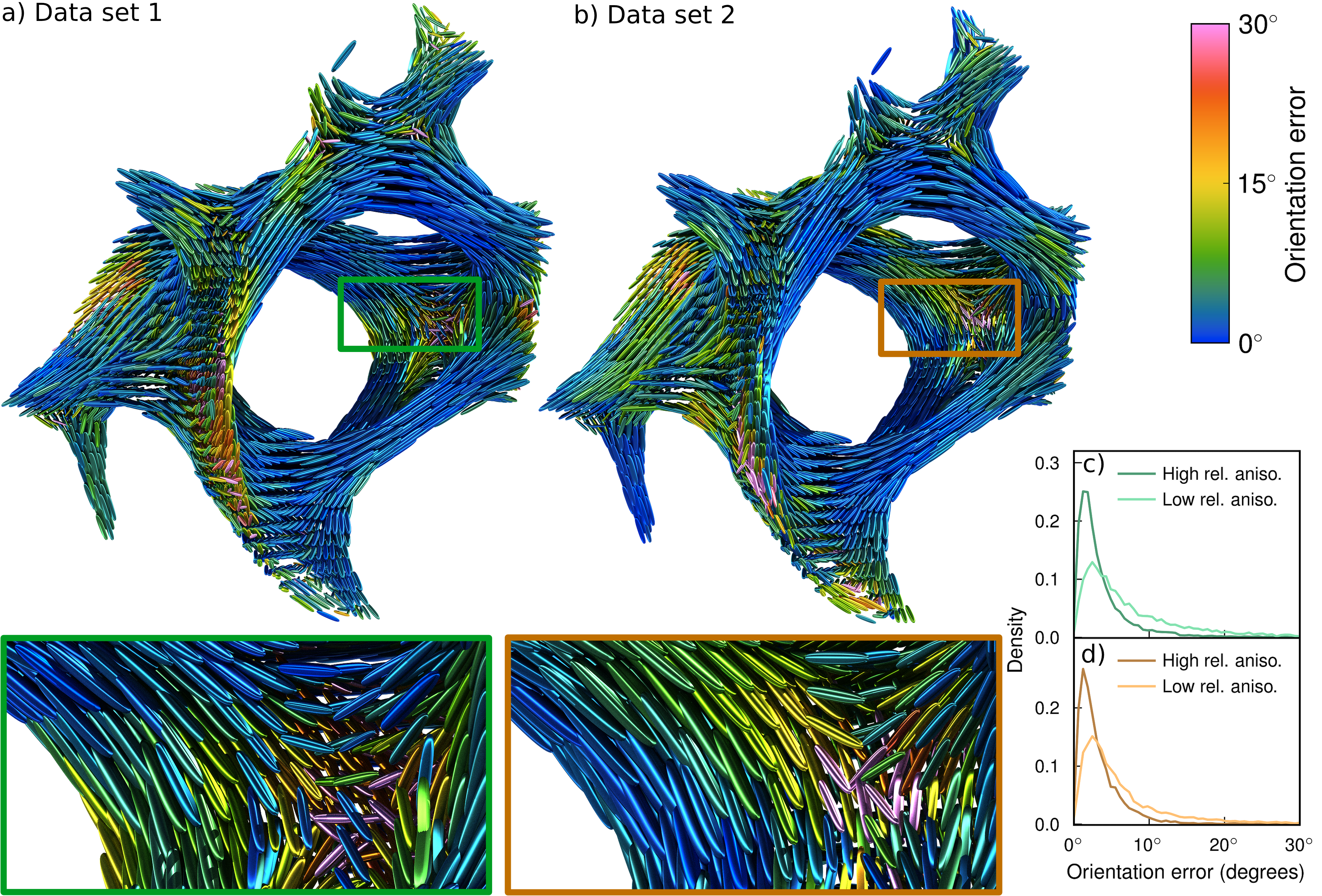}
  \caption{
    \textbf{Orientation errors.}
    \textbf{a)} Glyph render of orientations and errors of partial data set 1
    \textbf{b)} Glyph render of orientations and errors of partial data set 2
    \textbf{c)} Probability density plot of orientation errors of partial data set 1.
    \textbf{d)} Probability density plot of orientation errors of partial data set 2.
    The color of the glyph indicates the orientation error in that voxel compared to the full data set, with each glyph being scaled by the relative anisotropy in the partial reconstruction.
    The insets highlight an interface area where the different tendencies of the orientation errors for a) and b) can be seen, with a) showing larger errors for orientations closer to the $y$-axis, and b) showing larger errors orientations closer to the $x$-axis.
    The density plots show the orientation errors for high (greater than $0.6$) and low (less than $0.6$) relative anisotropy, showing that the error is greater in low relative anisotropy regions.
  }
  \label{fig:ori_error}
\end{figure}
One of the most important properties that can be retrieved from a \gls{saxstt} measurement is the local orientation, and it is therefore of interest to see how much uncertainty the missing wedge problem introduces in determining this.
\autoref{fig:ori_error} shows glyph renders of the orientation error for each partial reconstruction.
The orientation error is defined in \nameref{sect:methods}, using Eq.~\eqref{eq:great_circle_distance}.
Most orientations are determined to within an error of no more than \qty{10}{\degree}, as seen qualitatively for the blue and green colors in a), b) and quantitatively in the density plots c) and d).
The enlarged areas shown in green and orange rectangles show a region in the trabecular bone where differently oriented domains are intersecting.
This is the region where the orientation error is the largest in both partial data sets.
Comparing with \autoref{fig:scalars} d)--i), it can be seen that the larger orientation errors lie in regions where both the relative anisotropy and the fiber symmetry are small.
This means that the orientation is less well defined, and may include multiple orientations within a voxel.
This is consistent with the region containing an interface of domains of different orientation.
The same tendency is seen in the overall \gls{rsm} error which is largely similar to the distribution of the orientation errors (See \ref{sect:supp_errors}).
\begin{figure}
    \includegraphics[width=\linewidth]{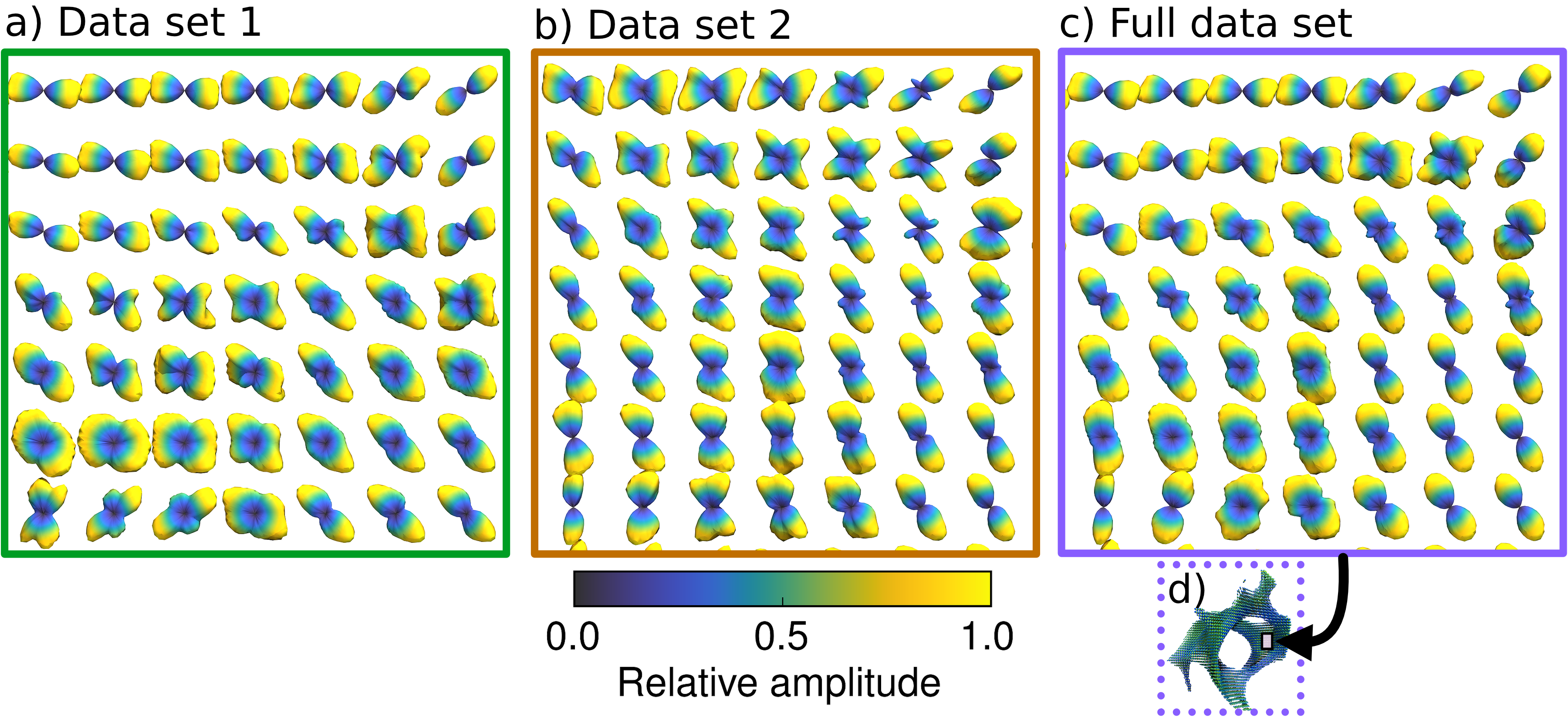}
  \caption{
    \textbf{Multiple orientations in interface region.}
    \textbf{a)} Data set 1, renders of Funk-Radon transform of anisotropic part of \gls{rsm}. 
    \textbf{b)} Data set 2.
    \textbf{c)} Full data set.
    \textbf{d)} Location of interface region in the sample.
    Maxima in the Funk-Radon transform indicate the orientation of each \gls{rsm}, and voxels with multiple local maxima appear to have multiple orientations.
    There are more apparent multi-orientation voxels in the partial reconstructions in a) and b), which is likely due to missing-wedge smearing of certain parts of the \gls{rsm} amplitude across real space.
  }
  \label{fig:multi_orientation}
\end{figure}
\autoref{fig:multi_orientation} illustrates a single slice from the enlarged region in \autoref{fig:ori_error} with larger errors, as well as low values of relative anisotropy and fibre symmetry.
The Funk-Radon transform of the \gls{rsm} shows that in this interface region multiple orientations are present inside single voxels.
The reconstruction with a model which does not impose strong symmetries, such as the grid of Gaussian radial basis functions used here, opens up the possibility to extract multiple orientations in each voxel.
However, comparing datasets 1 and 2, as well as the full data set illustrates that the missing wedge problem influences the accuracy of the reconstructed \gls{rsm} in the partial data set reconstructions.
The partial data set reconstructions in a) and b) have more voxels with apparent multi-orientation, and with a greater relative amplitude in the secondary orientation when compared to the full data set reconstruction in c).
This is likely due to missing-wedge smearing of certain parts of the \gls{rsm} amplitude across real space.
Thus, while multi-orientation analysis can be used to precisely localize this interface in a full-data reconstruction, the missing wedge problem makes this localization much more difficult in partial-data reconstructions.
\section{Conclusions}
In this work we have devised a scheme for complete acquisition of \gls{saxstt} data, and applied it to the analysis of a sample of trabecular bone.
Reconstructing incomplete as well as complete data sets and comparing them across both real and reciprocal space, we conclude that the understanding of data incompleteness in terms of the missing-wedge problem, as indicated by the computed quality factor, is consistent with the observed errors in the reconstruction.
Analyzing the orientations as well as scalar quantities, we find that the impact of the missing-wedge problem in a typical \gls{saxstt} analysis is limited, but appreciable in edge and interface areas.
In particular, the impact on mean \gls{rsm} amplitude and relative anisotropy is very limited, except for apparent artefacts in the mean.
Moreover, we observe that the impact of errors can be reduced by choosing the sample orientation during acquisition in a way that takes into account the missing wedge problem, i.e., by orienting the sample such that as much scattering as possible is close to the main axis of rotation.
Prior understanding of the nanostructure and expected \gls{rsm} of a sample, such as acquired by scanning \gls{saxs}, is crucial in this process.

This understanding could also be employed in various measures to reduce the impact of the missing wedge problem, e.g., by enforcing a particular \gls{rsm} symmetry.
Such symmetries can be encoded in the \gls{saxstt} basis set (as in \citeasnoun{liebi_aca_2018}, which used a spherical harmonic model that enforced rotational symmetry about an axis).
One disadvantage of encoding symmetries in the basis set is that more complex textures (such as the multi-orientation investigated in this work) cannot be captured.
However, symmetries can also be selectively enforced (based on a robust quantity, such as the relative anisotropy) in a post-processing step, or encouraged through regularization.
The further exploration of these possibilities and their impact on reconstruction quality is an interesting avenue for future research.
Finally, we remark that the complete acquisition scheme devised in this work is likely to be useful for specialized applications, such as the analysis of interface regions with overlapping domains of multiple orientations, or the reconstruction of especially complicated \glspl{rsm}.

\section{Data availability}
The data used in this work, along with code demonstrating the analysis and reconstructions which can be viewed in ParaView, is available via the DOI \href{https://doi.org/10.5281/zenodo.10995088}{10.5281/zenodo.10995088}.
\section{Acknowledgments}
This work was funded by the Swedish research council (VR 2018-041449) and the European research council (ERC-2020-StG 949301).
We acknowledge the Paul Scherrer Institut, Villigen, Switzerland for provision of synchrotron radiation beamtime at the beamline cSAXS of the SLS.
We extend our thanks to Manuel Guizar-Sicairos at the Paul Scherrer Institut and EPFL, for discussions on the experimental strategy to solve the missing wedge problem, as well as for co-authoring the experimental proposal.
We finally acknowledge Andreas Menzel at the cSAXS beamline at the Paul Scherrer Institut for assisting with the planning and administration of the experiment.
\section{Author contributions}
L.C.N.:  Conceptualization, writing, editing, coding, figure creation, theory, analysis. T.T.: Sample preparation, experiment, writing, figure creation, reviewing and editing. I.R.-F.: Experimental support, local contact, data curation, reviewing and editing. P.E.: Coding support, supervision, reviewing, and editing. M.L.: Conceptualization, methodology, supervision, experiment, analysis, reviewing, and editing.

\section{Competing interests}

The authors declare no competing financial or non-financial interest.


\referencelist{}

\end{document}


\title{
    Supplementary Information: \texorpdfstring{\\}{}
    Investigating the missing wedge problem in small angle x-ray scattering tensor tomography across real and reciprocal space
}

\author[a]{Leonard C.}{Nielsen}
\author[b,c]{Torne}{Tänzer}
\author[b]{Irene}{Rodriguez-Fernandez}
\author[a]{Paul}{Erhart}
\cauthor[a,b,c]{Marianne}{Liebi}{marianne.liebi@psi.ch}

\aff[a]{\chalmersphys}
\aff[b]{\affilpsi}
\aff[c]{\affilepfl}
\maketitle
\begin{description}
  \item[Supplementary Note 1] Real-space distribution of overall reciprocal-space errors
  \item[Supplementary Note 2] Gaussian kernel representation
  \item[Supplementary Note 3] Artefacts in data set 2
  \item[Supplementary Note 4] Quantitative analysis of scalar qualities
  \item[Supplementary Note 5] Solution algorithm
\end{description}

\section{Real-space distribution of overall reciprocal-space errors}
\label{sect:supp_errors}
\begin{figure}
    \includegraphics[width=\linewidth]{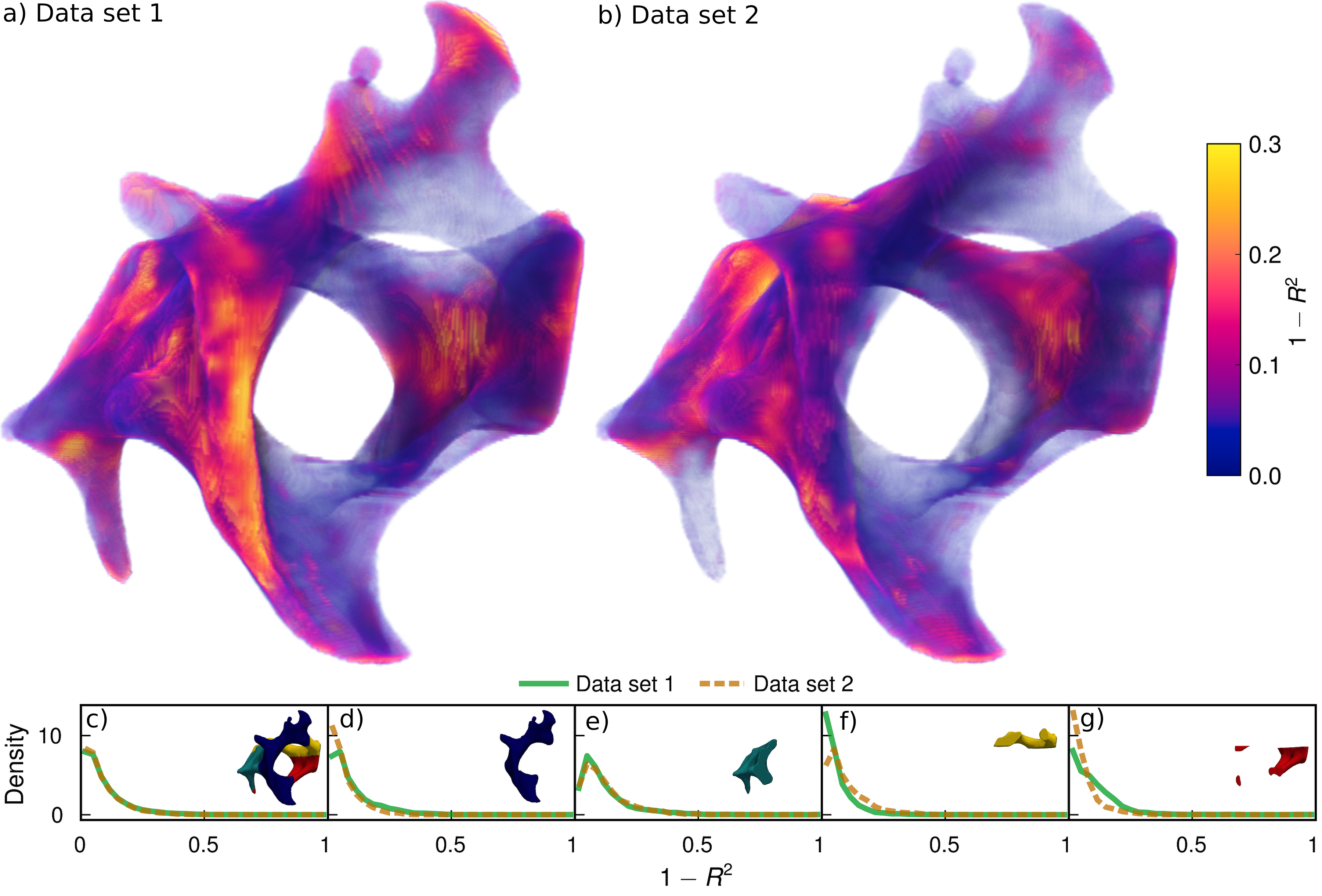}
  \caption{
    \textbf{Volume render of reconstruction error.}
    The volume render shows the spatial distribution of reconstruction errors between a reconstruction from the full dataset and a reconstruction from \textbf{a)} data set 1 only, and \textbf{b)} data set 2 only.
    \textbf{c)}--\textbf{g)} show the distribution of errors in the whole sample, as well as across 4 partitions, with each partition being indicated by a colored inset.
    The $R^2$ in the error is the squared Pearson Correlation Coefficient for the reciprocal space map in each voxel.
    The error distribution is similar to the distribution of orientation errors, with the highest errors occurring in interface and edge regions.
  }
  \label{fig:volume_error}
\end{figure}

\autoref{fig:volume_error} shows the overall reciprocal space map error.
Most of the errors appear to be in the flatter interface regions on the left- and right-hand side.
The distributions of the reciprocal space map errors are largely similar to the distributions of the orientation errors.
The distributions of errors across 4 partitions show that the better-performing data set varies depending on the local structure.
Across the whole sample, panel c), the two data sets perform very similarly.
The blue area, panel d), has several protrusions which generally appear to perform better in data set 2, which therefore performs somewhat better overall.
In the teal region, panel e), which is more blocky in its structure, the data sets perform approximately equally, but as is evident from the renders in panels a) and b), the errors are differently distributed.
In the largely horizontally-oriented yellow part, panel f), data set 1 performs better, and the slightly more vertically-oriented red part, panel g), has data set 2 performing better.
This shows the importance of considering the overall sample structure in order to optimize data acquisition.
The error distributions are similar to the correlation coefficients obtained for simulated data in \citeasnoun{nielsen_tt_2023}.

\section{Gaussian kernel representation}
\label{sect:gaussian_kernels}
\begin{figure}
    \includegraphics[width=\linewidth]{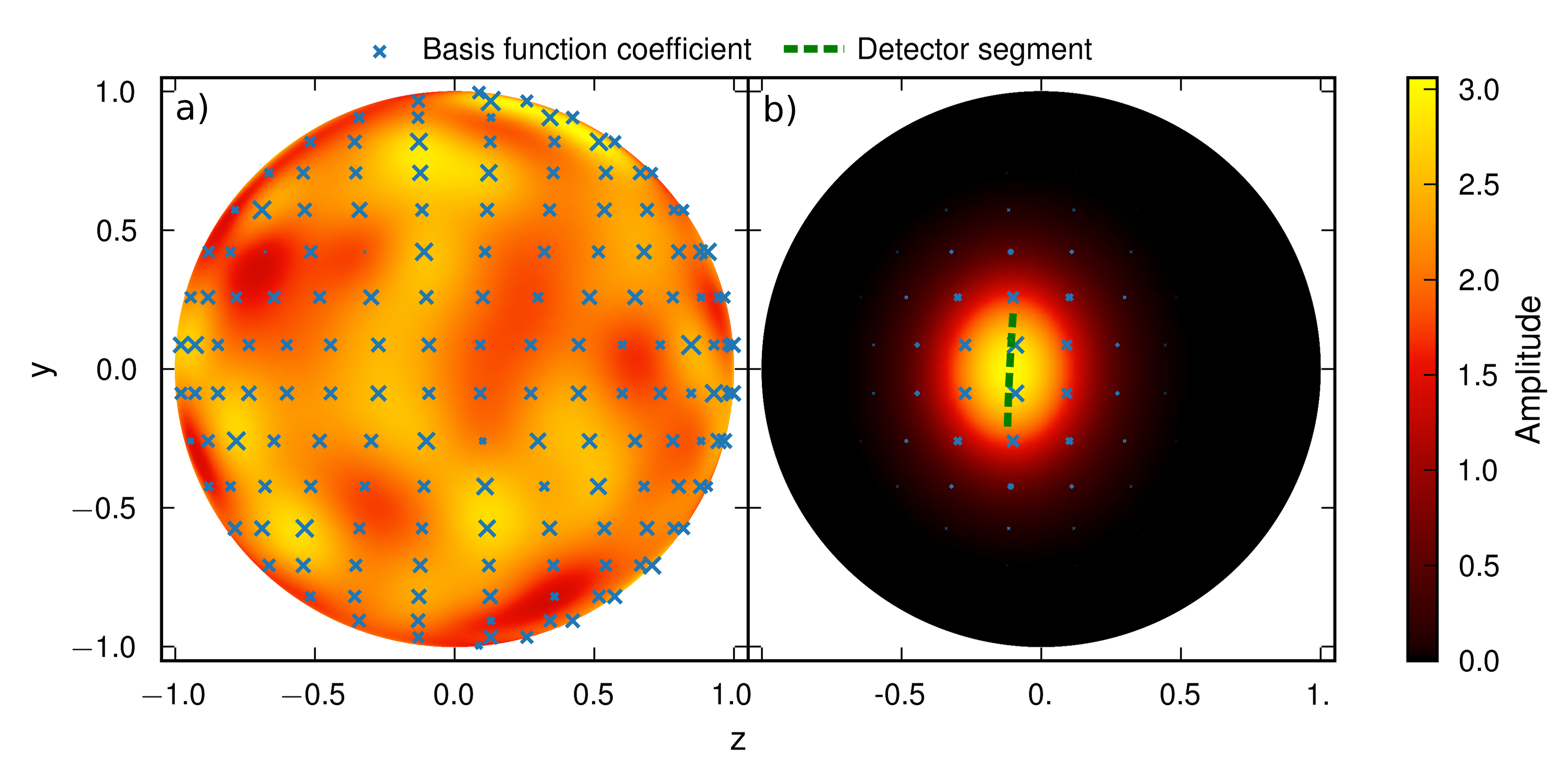}
  \caption{
    \textbf{Gaussian kernel representations of spherical functions.}
    \textbf{a)} Spherical function with white noise coefficients.
    \textbf{b)} Mapping from Gaussian kernels to detector segment.
    The blue crosses show the location and magnitude of each coefficient.
    The green line shows the location of a detector segment on the sphere.
    
  }
  \label{fig:gk_plot}
\end{figure}
In \autoref{fig:gk_plot} we see two illustrations of the Gaussian kernel representation, including how the kernel functions are distributed using a modified Kurihara mesh, what a function consisting of noise looks like, and how the detector segments map to the sphere \cite{kurihara_mwr_1965}.
The principal modification of the mesh consists in enforcing Friedel symmetry.
In addition, the mesh is rearranged to make the distribution of points somewhat more regular.
Moreover, the effect of the anisotropic distribution of points on the sphere is reduced by scaling each basis function by the value that is projected onto the point it is centered on in the grid when all coefficients are unity; this is the so-called ``auto-projection'' mentioned in the main work.
Thus, basis functions centered on points which are in denser areas of the grid are scaled down, and functions centered on points in less dense areas of the grid are scaled up.
At the equator, the Kurihara mesh contains only a semicircle of basis functions, since the representation respects Friedel symmetry.
The number of points in the grid is decreased in each circle with increasing radius, to maintain approximately constant density of points.
The standard deviation of the kernels is set based on the approximate distance between points to be wide enough to allow for the easy representation of unimodal distributions by multiple points.
In addition, as can be seen in the b) panel, the width of the kernels implicitly enforce some of the continuity which is assumed of the reciprocal space map in the reconstruction.

\section{Artefacts in data set 2}
\label{sect:supp_artefacts}
\begin{figure}
    \includegraphics[width=\linewidth]{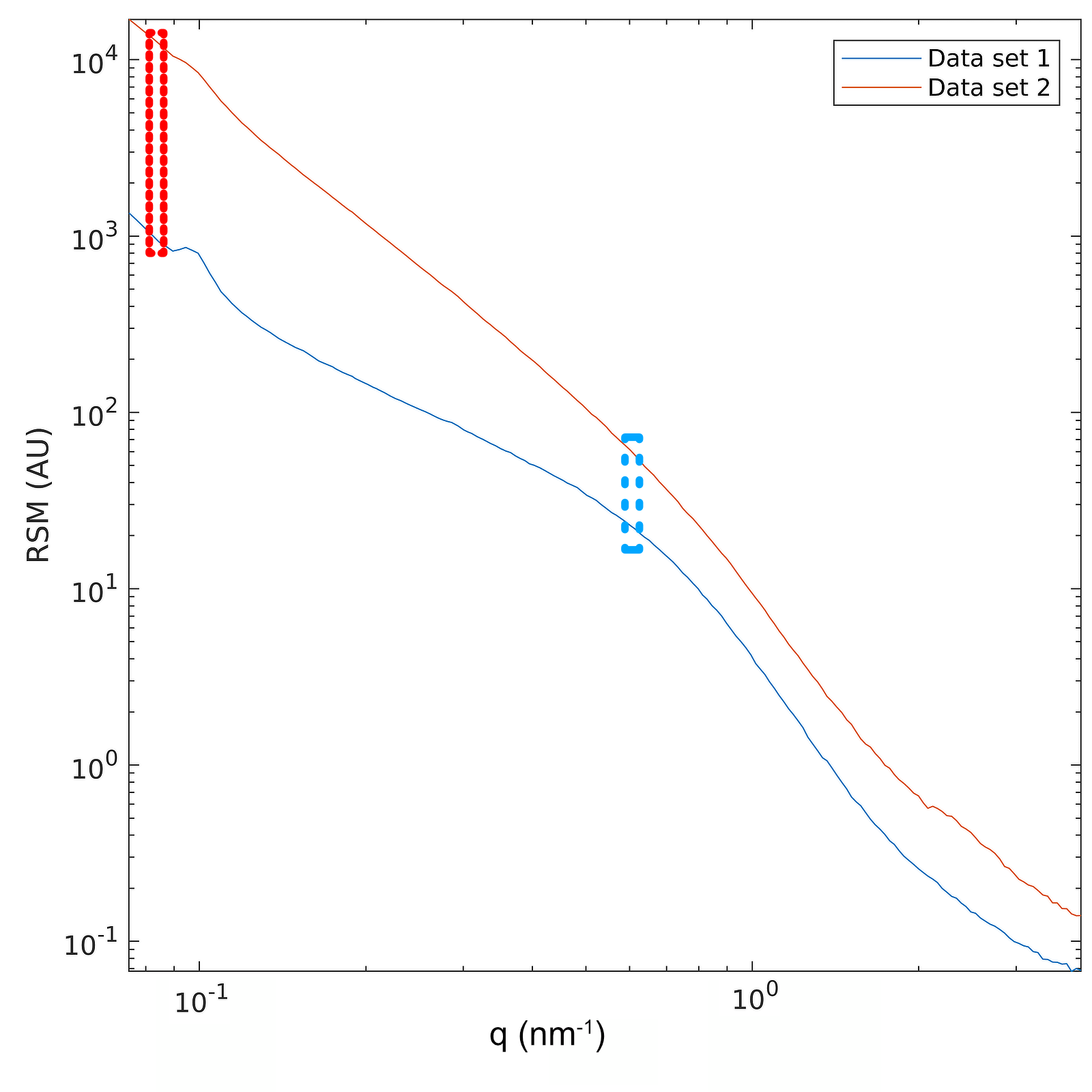}
  \caption{
    \textbf{Scattering curves of data sets 1 and 2 for one pixel}
    Note the difference both in magnitude and shape of the scattering curve between the two data sets.
    The area enclosed by the red dashed outline indicate the q-region chosen for illustrating the large influence of the artefacts.
    The area enclosed by the blue outline indicates the q-region used for the reconstruction in the main work.
    }
  \label{fig:scattering_curve}
\end{figure}
\begin{figure}
    \includegraphics[width=\linewidth]{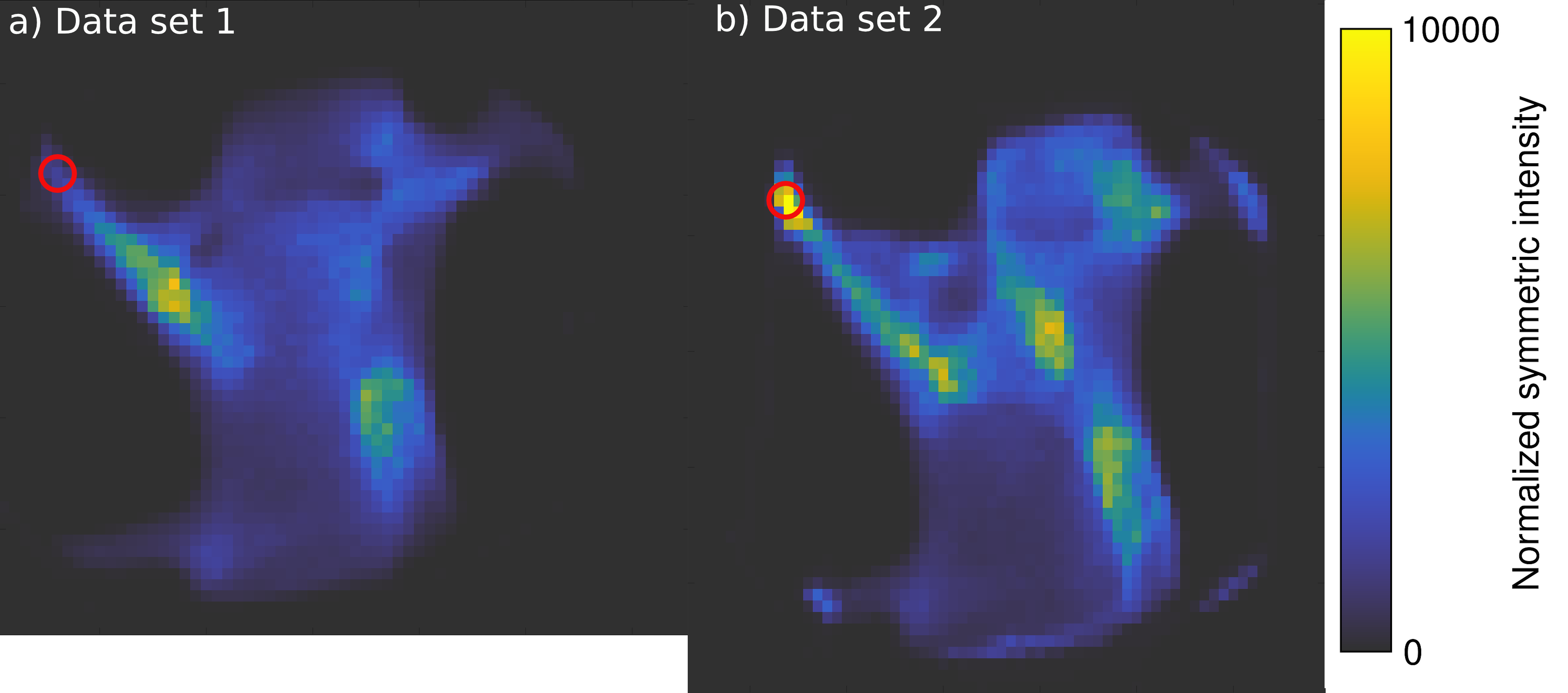}
  \caption{
    \textbf{Raster scans from data sets 1 and 2}
    \textbf{a)} Raster scan from data set 1.
    \textbf{b)} Raster scan from data set 2.
    The raster scans are colored by the transmission-normalized symmetric intensity around the $q$-value \qty{0.0845}{\per\nano\meter}.
    The red circle indicate the pixel of the scattering curve shown in \autoref{fig:scattering_curve}.
    }
  \label{fig:raster_scans}
\end{figure}

In \autoref{fig:scattering_curve} the azimuthally integrated scattering curve from a single measurement is shown for each of the data sets.
The chosen points measure approximately the same location in the sample from approximately the same angle.
The red outline highlights a high-q region shown in the raster scans in \autoref{fig:raster_scans}; the q-value for these scans is \qty{0.0845}{\per\nano\meter}.
The areas circled in red are the points where the scattering curves in \autoref{fig:scattering_curve} were measured.
As can be seen, the symmetric intensity is higher by an order of magnitude in data set 2, even though the views are very similar.
The difference in the curves is much less around the higher q-range of \qtyrange{0.597}{0.607}{\per\nano\meter}, which is used in the main work and highlighted in green in \autoref{fig:scattering_curve}, although still significant for the selected point.

\begin{figure}
    \includegraphics[width=\linewidth]{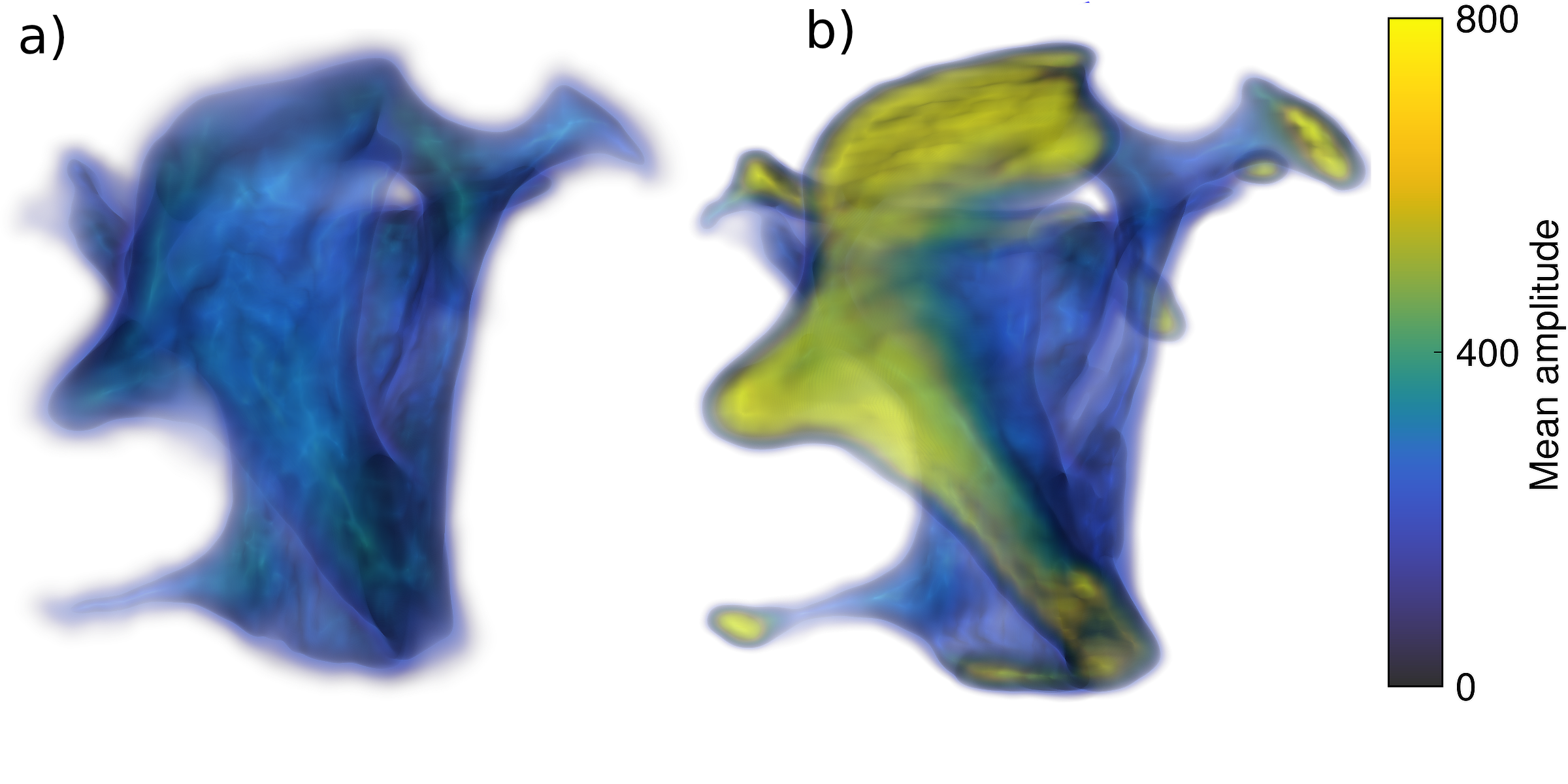}
  \caption{
    \textbf{Volume renders of mean amplitude of low q-range reconstruction.}
    \textbf{a)} Volume render from data set 1.
    \textbf{b)} Volume render from data set 2.
    Note that the values above 800 are given zero opacity in order to render the figure clearly.
  }
  \label{fig:volrend_diff_sym}
\end{figure}

In \autoref{fig:volrend_diff_sym} we can see the differences in the mean amplitude (the reconstruction's counterpart to the normalized symmetric intensity) between the two data sets at this lower q-range in many regions, especially those close to the edges.
The differences both in the raster scan and in this reconstruction is consistent with some artefact being introduced between the two measurements, with the most likely cause being the water-soluble glue mentioned in the main work.

\section{Quantitative analysis of scalar quantities}
\label{sect:supp_abs_dev}

\begin{figure}
    \includegraphics[width=\linewidth]{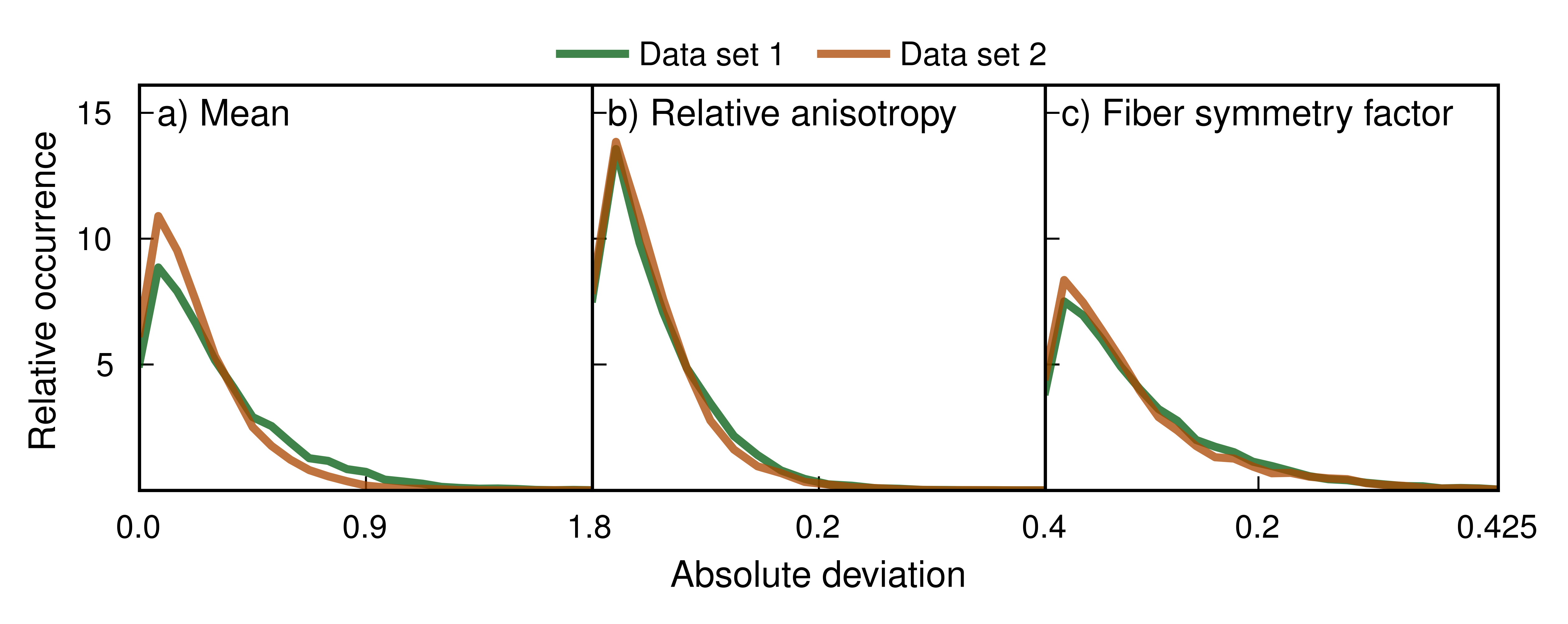}
  \caption{
    \textbf{Absolute deviations of scalar quantities in sample.}
    \textbf{a)} Absolute deviation of mean.
    \textbf{b)} Absolute deviation of relative anisotropy.
    \textbf{c)} Absolute deviation of symmetry factor.
  }
  \label{fig:absolute_deviations}
\end{figure}

Frequency plots of the absolute deviations between the full reconstruction and each partial reconstruction for three scalar quantities are shown in \autoref{fig:absolute_deviations}.
The limits for the plots are set based on histograms of the quantities.
This shows the relative anisotropy as the most robust quantity.
The mean is likely more robust than suggested here, due to artefacts introduced during re-mounting.

\begin{figure}
    \includegraphics[width=\linewidth]{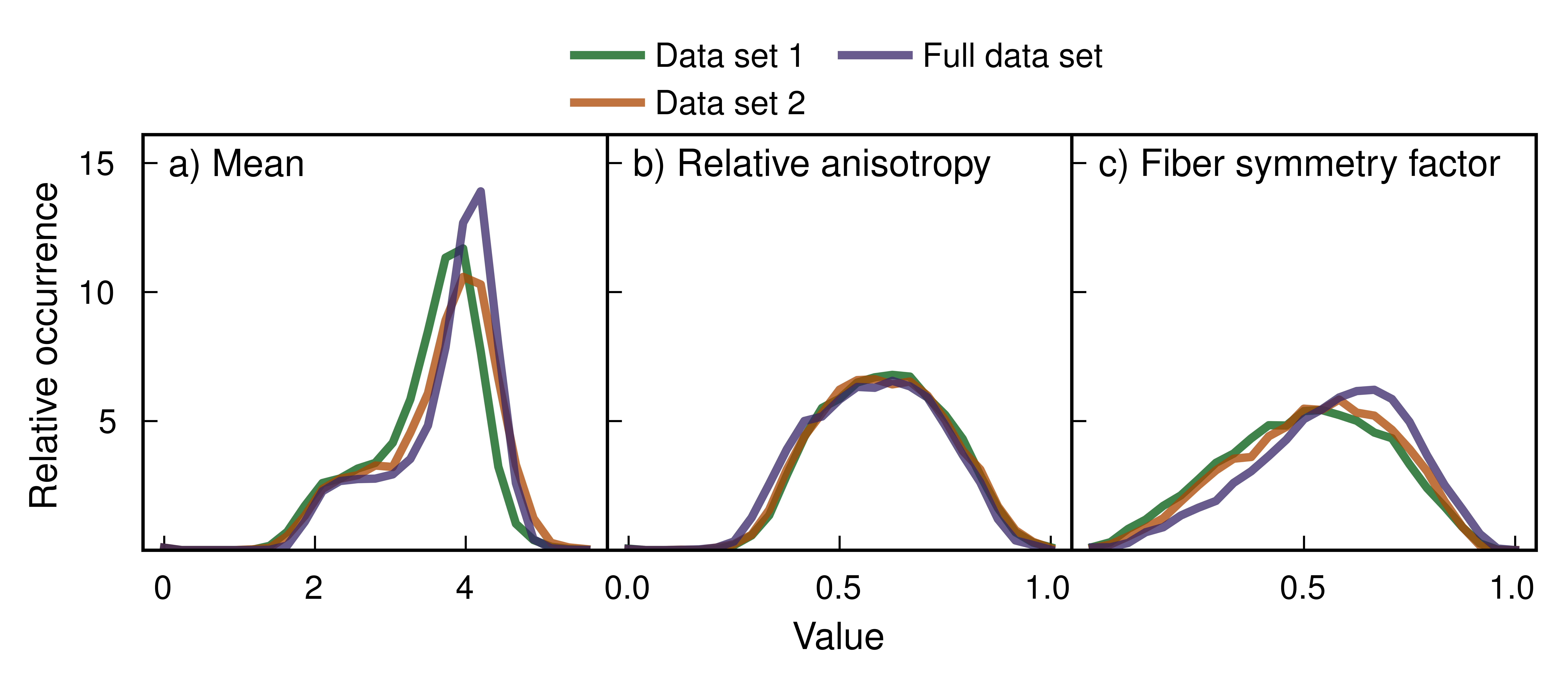}
  \caption{
    \textbf{Relative occurrence of scalar quantities in sample.}
    \textbf{a)} Relative occurrence of mean.
    \textbf{b)} Relative occurrence of relative anisotropy.
    \textbf{c)} Relative occurrence of symmetry factor.
  }
  \label{fig:relative_occ}
\end{figure}

In \autoref{fig:relative_occ} we see the distribution of the three scalar quantities in the sample.
It is clear from this plot that the mean of data set 2 is shifted in overall magnitude, which would be reflected in overall greater deviations in \autoref{fig:absolute_deviations} which reflect experimental artefacts, rather than theoretical robustness.

\section{Optimization}
\label{supp:optimization}

\referencelist{}